\documentclass[12pt,preprint]{aastex}

\shorttitle{Radio-loud NLS1 galaxies}
\shortauthors{Komossa et al.}

\begin{document}


\title{Radio-loud Narrow-Line Type 1 Quasars} 

\author{Stefanie Komossa, Wolfgang Voges}  
\affil{Max-Planck-Institut f\"ur extraterrestrische Physik,
    Postfach 1312, 85741 Garching, Germany; skomossa@mpe.mpg.de}

\and

\author{Dawei Xu} 
\affil{National Astronomical Observatories, Chinese Academy of Sciences, 
A20 Datun Road, Chaoyang District, 
Beijing 100012, China}

\and

\author{Smita Mathur}
\affil{Department of Astronomy, The Ohio State University, 140 West 18th Avenue, 
Columbus, OH 43210, USA}

\and

\author{Hans-Martin Adorf}
\affil{Max-Planck-Institut f\"ur extraterrestrische Physik,
    Postfach 1312, 85741 Garching, Germany;
 and  
Max-Planck-Institut f\"ur Astrophysik, Karl-Schwarzschild-Strasse 1,
85748 Garching, Germany}

\and 

\author{Gerard Lemson}
\affil{Max-Planck-Institut f\"ur extraterrestrische Physik,
 Postfach 1312, 85741 Garching, Germany}

\and 

\author{Wolfgang J. Duschl}
\affil{Institut f\"ur Theoretische Astrophysik, Albert-Ueberle-Str. 2,
   69120 Heidelberg Germany;
 and
Steward Observatory, The University of Arizona, 933 N. Cherry Ave.,
   Tucson, AZ 85721, USA}

\and

\author{Dirk Grupe}
\affil{Astronomy Department, Pennsylvania State University, 525 Davey Lab, 
University Park, PA 16802, USA} 

\begin{abstract}

We present the first systematic study of (non-radio-selected) radio-loud 
narrow-line Seyfert 1 (NLS1) galaxies. 
Cross-correlation of
the `Catalogue of Quasars and Active Nuclei' with several radio and optical catalogues
led to the identification of 
 $\sim$11 radio-loud NLS1 candidates including 4 previously known ones. 
This study almost triples the number of known radio-loud NLS1 galaxies if all candidates
are confirmed. 
Most of the radio-loud NLS1s are compact, steep spectrum sources 
accreting close to, or above, the Eddington limit. 
The radio-loud NLS1s of our sample are remarkable in that they
occupy a previously rarely populated regime in NLS1 multi-wavelength
parameter space.
While their [OIII]/H$\beta$
and FeII/H$\beta$ intensity ratios almost cover the whole range
observed in NLS1 galaxies, their radio properties extend 
the range of radio-loud objects to those with small widths of the broad Balmer lines.   
Their black hole
masses 
are generally at the upper observed end  
among NLS1 galaxies, but 
are still
unusually small in view of the radio loudness of the sources.  
Among the radio-detected NLS1 galaxies, the radio index $R$
 distributes quite smoothly up to the critical value of $R \simeq 10$
and covers about 4 orders of magnitude in total.
Statistics show that $\sim$7\% of
the NLS1 galaxies 
are formally radio-loud
while only 2.5\% exceed a radio index $R > 100$. 
Implications for NLS1 models are discussed. 
Several mechanisms
are considered as explanations for 
the radio loudness of the NLS1 galaxies 
and for the lower frequency of radio-louds 
among NLS1s than quasars.  
While properties of most sources (with 2-3 exceptions) 
generally do not favor relativistic beaming,
the combination of accretion mode and spin may explain the
observations. 
\end{abstract}


\section{Introduction}

The radio-loud radio-quiet bimodality of quasars is 
one of the long-standing unsolved problems 
in the research of Active Galactic Nuclei (AGN).
While the presence of two classes of quasars,
radio-louds and radio-quiets with a deficiency of
sources at intermediate radio powers, was historically   
apparently well established (e.g., Kellerman et al. 1989,
Miller et al. 1990, Visnovsky et al. 1992), 
other studies have questioned the mere existence of a radio dichotomy 
(e.g., White et al. 2000, Cirasuolo et al. 2003), 
 or have pointed out that larger samples are needed to put
the bimodality on a firmer statistical basis (Hooper et al. 1995). 
On the other hand  it was argued that 
carefully selected small, well-studied samples still reveal the bimodality
(e.g., Sulentic et al. 2003), and that
it is still present in a large sample of SDSS-FIRST galaxies  
(Iveciz et al. 2002).  Approximately 15\% of the quasar population 
 is radio loud (Urry \& Padovani 1995) for reasons that are still unknown.   

A second question that is presently under close scrutiny is  
whether or not there are relations
between black hole (BH) mass and radio loudness, and especially,  whether
or not there is a limiting BH mass above 
which objects are preferentially radio loud, and
whether or not radio-louds and radio-quiets show the same spread
in their black hole masses  (e.g., Laor 2000,  Lacy et al. 2001,
Oshlack et al. 2002, Woo and Urry 2002, Shields et al. 2003, McLure \& Jarvis 2004,
Metcalf \& Magliocchetti 2006, Liu et al. 2006).

While most attention so far concentrated on the 
radio-loud radio-quiet distinction
of quasars, extending the studies
to other subtypes of AGN may shed new light on both, the existence
of a radio-loud radio-quiet bimodality and on the cause of radio loudness.
We have started an investigation of the radio properties
of Narrow-line Seyfert 1 galaxies.  
Apart from the
topic just mentioned, the analysis 
of their radio properties is also of interest for addressing 
open questions in the study of NLS1s, as detailed below.  

NLS1 galaxies
were identified by Osterbrock \& Pogge (1985) as objects 
with small widths of the broad Balmer lines. These go hand in hand with 
weak [OIII]5007/H$\beta_{\rm totl}$ emission
and strong emission from FeII complexes (e.g., Boroson \& Green 1992). 
Physical drivers of the NLS1 phenomenon and correlations among emission-line
and continuum properties
are not
yet well understood. There is growing evidence, though, that most NLS1s 
are objects with high accretion rates, close to or even super-Eddington,     
and low black hole masses 
(e.g., Boroson \& Green 1992, Pounds et al. 1995, Wang et al. 1996, 
Boller et al. 1996, Laor et al. 1997, 
Czerny et al. 1997,
Marziani et al. 2001, Boroson 2002, Simkin \& Roychowdhury 2003,
Xu et al. 2003, Kawaguchi 2003, Wang \& Netzer 2003, Grupe 2004,
Grupe \& Mathur 2004, Botte et al. 2004, Collin \& Kawaguchi 2004). 
Apart from that,
several
other parameters were suggested to 
influence 
the main properties of NLS1 galaxies, including 
orientation 
(e.g., Osterbrock \& Pogge 1985, Bian \& Zhao 2004), 
winds and density effects (e.g., Lawrence et al. 1997, Wills et al. 2000,
Bachev et al. 2004, Xu et al. 2003, 2006),
metallicity (e.g. Mathur 2000, Komossa \& Mathur 2001, 
Shemmer \& Netzer 2002, Nagao et al. 2002, Warner et al. 2004, Fields et al. 2005),
and 
absorption (e.g., Komossa \& Meerschweinchen 2000,
Gierlinski \& Done 2004).   
There is evidence that NLS1s follow a different  
black hole mass -- velocity dispersion ($M_{\rm BH}-\sigma$)
relation than broad line Seyfert 1 galaxies
(e.g., Mathur 2001, Grupe \& Mathur 2004, 
Mathur \& Grupe 2005)  
with important implications for galaxy evolution and black hole
growth.

While the X-ray and optical properties of
NLS1s were explored intensively in the last decade, 
relatively little is known about their radio properties. 
Ulvestad et al. (1995) observed 7 NLS1 galaxies with the VLA
and concluded that they are only of modest radio power 
and that radio emission is more compact than a few hundred parsec. 
Moran (2000) studied 24 NLS1s
with the VLA and found that 
most of the sources are unresolved and show relatively steep spectra.   
Stepanian et al. (2003) found 9 radio(FIRST)-detected NLS1s among 
26 NLS1s of the Second Byurakan
Survey, all of them radio-quiet.
Greene et al. (2006) performed radio observations of 19 galaxies
with low BH masses, several of them with optical spectra
similar to NLS1s. They only detected one source.  

So far, only a few radio-loud NLS1s
have been identified{\footnote{there
are preliminary reports that optical identifications
of FIRST radio sources produce a higher rate of radio-loud NLS1s
than optical selection (Whalen et al. 2001; see 
our Section 5.8 for a comparison with Whalen et al. 2006)}}. 
PKS0558-504 (Remillard et al. 1986, Siebert et al. 1999),
RXJ0134-4258 (Grupe et al. 2000), 
SDSSJ094857.3+002225 (Zhou et al. 2003) 
and SDSSJ172206.03+565451.6 (Komossa et al. 2006)
all exceed radio indices $R$=10 and show optical NLS1 spectra. 
RGB J0044+193 (Siebert et al. 1999)
is radio-quiet most of the time (Maccarone et al. 2005), 
but may have been radio-loud at
the epoch of the 87GB survey (Siebert et al. 1999, Maccarone et al. 2005)
due to variability in the radio band.
PKS 2004-447 (Oshlack et al. 2001)
is a (non-typical) NLS1 or possibly
a narrow-line radio galaxy (Zhou et al. 2003)
and is very radio-loud.

It is still a puzzle why radio-loud NLS1 galaxies are scarce.   
In particular, it is still unclear whether they are truly preferentially radio quiet,
or rather the scarcity of radio-louds is due to some selection effect. 
This question can be addressed by a systematic search for, and study of,
radio-loud NLS1s. 

The study of radio properties of NLS1s, in particular their radio loudness
and radio variability,  
also allows us to re-address the question whether NLS1 galaxies are preferentially
viewed face on. 
Furthermore, radio observations of NLS1 galaxies 
 provide constraints on the coupling between jets and accretion disks
(e.g., Zdziarski et al. 2003).
There are indications that objects with accretion rates
close to the Eddington rate
tend to be radio-weaker; a behavior observed in Galactic X-ray binaries
in soft/high-state
and AGN (Maccarone et al. 2003, Greene et al. 2006, and references therein).    

Here, we report results from a search for radio-loud and `almost radio-loud'
NLS1 galaxies,
based on cross-correlating a catalogue of optically identified NLS1s
with radio catalogues. 
This is the first systematic study of (non-radio-selected) 
radio-loud NLS1 galaxies, and
of their radio, optical and X-ray properties. 
We use the term ``Narrow-line Seyfert 1 (NLS1) galaxy'' collectively
for high-luminosity and low-luminosity objects, i.e. for
narrow-line type 1 quasars and for narrow-line Seyfert type 1 galaxies. 
A cosmology with $H_{\rm 0}$=70 km/s/Mpc, $\Omega_{\rm M}$=0.3
and $\Omega_{\rm \Lambda}$=0.7 is adopted throughout this paper. 
 
This paper is organized as follows:
In Section 2 we describe our methods for cross-matching
a large number of catalogues and present candidate radio-loud
NLS1 galaxies. In Sections 3 and 4 we provide results from our analysis of
the optical and X-ray spectra of these NLS1s.  
Consequences of our results for NLS1 models, and
mechanisms to explain the lower frequency of radio-louds
among NLS1s than quasars are discussed in Section 5.

\section{GAVO search for radio-loud NLS1 galaxies}

\subsection{Search methods} 

In order to search for radio-loud 
NLS1 galaxies and study their properties, we extracted all
NLS1 galaxies included in the 11th edition of
the  ``Catalogue of Quasars and Active Nuclei''
compiled by Veron-Cetty \& Veron (2003;   
henceforth referred to as VQC), separately for the catalogue
of ``faint'' objects (i.e., Seyferts fainter than absolute magnitude 
$M=23$) and
the ``bright'' ones (i.e, quasars brighter than $M=23$).
We present results for the narrow-line type 1 quasars here, while 
results for the Seyfert galaxies will be reported elsewhere. 
The defining criterion for inclusion as NLS1 galaxy
in the VQC is a width of the H$\beta$ emission line
FWHM$_{\rm H\beta} < 2000$ km/s. 
The list of the 128 selected narrow-line quasars 
was then cross-correlated
with the  
FIRST (``Faint Images of the Radio Sky at Twenty centimeters'', at 1.4 GHz; 
  White et al. 1997, Becker et al. 1995),
NVSS (``NRAO VLA Sky Survey'', at 1.4 GHz; Condon et al. 1998),
SUMSS (``Sydney University Molonglo Sky Survey'', at 843 MHz; Mauch et al. 2003),
WENSS (``Westerbork Northern Sky Survey'', at 0.33 GHz; de Bruyn et al. 1998),
PMN (``Parkes-MIT-NRAO'' radio survey including all, the  
Southern Survey, Zenith Survey, Tropical Survey, and Equatorial Survey, at 4.85 GHz; Griffith et al. 1994),
87GB (``Green Bank Radio Survey'', at 4.85 GHz; Gregory \& Condon 1991)
and the PKS (``Parkes Radio Survey'', at 2.7 GHz; Wright \& Otrupcek 1990) 
radio catalogues, in order to select all radio-detected
NLS1s, and among these the radio-loud ones. 
The input list was simultaneously cross-correlated with the 
USNO-B1 (Monet et al. 2003), USNO-A2 and GSC2.2 optical photometry catalogues
in order to obtain blue magnitudes. 

For cross-matching of radio catalogues with optical
catalogues we used the cross-matcher application developed 
within the German
Astrophysical Virtual Observatory (GAVO){\footnote{See http://www.g-vo.org/}} project.
The GAVO cross-matcher is a multi-archive, multi-server,
multi-catalogue, statistical spatial matcher operating
solely on astrometric coordinates and their uncertainties
(Adorf et al. 2003, 2005). 

In our case, it takes as input an extract of the `Veron quasar catalogue' 
which includes all NLS1 galaxies in that catalogue and
their coordinates.  A query module then accesses 
the VizieR archive{\footnote{See http://vizier.u-strasbg.fr/viz-bin/VizieR}} 
and extracts coordinates and flux and magnitude data from selected catalogues
in the archive. 

While the VizieR match-list services are capable of deterministically matching
the input list against a single catalogue,
we then proceed in matching the different output lists against each other
in order to find the most likely single multiwavelength ``counterpart''. 
In a first step, 
match-candidates are generated 
in the following way: combinations of potential counterparts are formed, which
are viewed as hypothetical ``clusters'' on the sky. For each such
cluster its central position is computed, followed by a computation
of the projected distances on the sky between each counterpart candidate and
the central cluster position.

Next, based on the astrometric positions and associated
uncertainties, for each counterpart in a given match candidate
cluster its isotropic Mahalanobis distance (Mahalanobis 1936, Devroye et al. 1996)
to the central cluster position is
computed. It measures the statistical distance to the cluster
center.
For each cluster a goodness-of-fit criterion is computed, namely the
standard~$\chi^2$-metric, which is simply the sum of the squared
Mahalanobis distances. It can be viewed as a measure of compactness
of the cluster, where instead of the physical distances, the
statistical distances are being used.

Since the number of counterparts contributing to a match candidate
generally differs from candidate to candidate, we cannot use the~$\chi^2$
metric directly for comparing the goodness of the match. Instead the  
reduced~$\chi^2$ metric is 
computed.  
For the required degrees
of freedom~$N_{\rm f}$ we use
$ N_{\rm f} = 2 N_{\rm i} - 2$, 
where~$N_{\rm i}$ denotes the number of counterparts contributing to a
match candidate. We subtract~$2$ in order to allow for the
estimation (``fitting'') of the two sky-coordinates, i.e.\ right
ascension and declination, when computing the cluster center.
A reduced-$\chi^2$ threshold 
parameter is used for discriminating
against unreasonable match combinations.
Finally, for each remaining match candidate, the cross-matcher
extracts the astrometry and optical and radio photometry data.    

Position uncertainties, inherent to the different (radio) surveys, 
are either taken into account on 
an object-by-object basis if individual position errors are 
available (this is generally the case for the FIRST and 87GB catalogues in VizieR; 
typically 1$^{\prime\prime}$ (FIRST) and $<$ 30$^{\prime\prime}$ (87GB))
or else typical average position uncertainties $d$ for the different
catalogues are used (explicitly, $d$=1$^{\prime\prime}$ (USNO-A2,USNO-B1,GSC2.2),
7$^{\prime\prime}$ (NVSS), 30$^{\prime\prime}$ (PKS and PMN), 10$^{\prime\prime}$ (SUMSS and WENSS)). 
The catalogue-specific search radii were typically larger by a factor of 3.

\subsection{Safety checks and data screening}

While this automated way of simultaneously cross-correlating various catalogues 
and of selecting counterparts is generally very successful,  there are still
a number of ``peculiarities'' inherent to the way the catalogues
were produced in general, or to individual galaxies in particular,
which affect the cross-matching output and which
make it necessary to check a number of results
``by hand''.  While the long term goal is to include most
of these effects into the cross-correlation software, in this pilot
project we paid special attention to actually identifying potential
problems and effects that need closer attention in the future when
applying similar cross-correlation methods to much larger samples. 

Nearby bright galaxies are missed in blue magnitude catalogues because
they are too extended and too bright for standard automated magnitude measurements 
to still work. This is not a problem for the present sample 
since the quasars are sufficiently distant but will be more severe for
samples of nearby Seyfert galaxies.  

A few catalogued NLS1s still have optical
coordinate off-sets which are larger than expected in
the optical band. One reason is that their IRAS coordinates rather than
optical coordinates were still listed, which come with a larger uncertainty.
Therefore, counterparts in blue magnitudes and FIRST radio sources
might be missed.  For the present sample, IRAS11598-0112 and RXJ01354-0426
were not assigned optical counterparts by the original
cross-correlation procedure. IRAS11598-0112 had its
IRAS coordinates catalogued, while the true optical coordinate then turns out
to be consistent with the FIRST radio position. The optical counterpart 
of RXJ01354-0426 is off-set by $\sim$3-4$^{\prime\prime}$ from the 
coordinate in the VQC. However, since there
is no other good multi-wavelength counterpart in the field, we assume
that optical, radio and X-ray source are one and the same.

Bright radio sources are often detected in several surveys with widely different
positional uncertainties. 
In that case it is generally
a  good idea to then use the smallest positional error among the
radio catalogues for further cross-correlation with non-radio catalogues
to reduce the number of potential multi-wavelength match candidates. 
In the present study, in two cases, a bright source was 
detected in several radio surveys, 
suggesting we deal with one and the same source, but
the source coordinates in those radio surveys  with small coordinate
uncertainties were no longer consistent with the optical position.     
In the specific case of RXJ07101+5002, only a 87GB counterpart was  
found by cross-correlation which is 45$^{\prime\prime}$ off the
optical position. The 87GB source has counterparts in several 
other radio catalogues including NVSS and FIRST, which
show the same off-set from the optical coordinates and similar
radio fluxes. This means 
the radio source is one and the same in all radio catalogues. However,
given the small coordinate uncertainty in the FIRST catalogue, it
can no longer be related to the optical NLS1 galaxy. The FIRST
image reveals a complex radio structure with several bright knots
and lobes. We have checked that none of these sub-structures
coincides with the optical NLS1. Rather, the core of the radio
galaxy does have a `normal' galaxy as optical counterpart, reported by Bauer et al.
(2000). The question remains whether the {\em X-ray source} RXJ07101+5002 then
actually is the counterpart to the radio source (as often assumed;
e.g., Bauer et al. 2000), or whether it is the counterpart to the
optical NLS1 galaxy, as concluded by Xu et al. (2003) who provided
the optical spectroscopy of that source. The radio galaxy is radio loud 
with $R_{\rm 1.4}$=64. We exclude it from the sample, though, since 
the NLS1 galaxy is not its counterpart.    
In the case of SBS1330+519, only a WENSS counterpart was found,
off-set by 20$^{\prime\prime}$. The WENSS source has a FIRST and NVSS
counterpart which, however, are too far off from the optical
position of SBS1330+519 to be still considered as counterparts. 
SBS1330+519 was thus
excluded from our sample. We note in passing that the relatively
strong radio source itself does have a very faint optical counterpart
in the Sloan Digital Sky Survey (SDSS; York et al. 2000) photometric 
catalogue (Abazajian et al. 2005), SDSSJ133219.38+514439.9.
With SDSS $r$ and $g$ magnitudes of 21.5 and 22.3 mag, respectively, this background
object is very radio loud ($R_{1.4}$=8940). No optical spectrum is available.  

Radio images of all radio-detected NLS1s were inspected by eye to 
convince ourselves of the reliability of counterpart identification. 

\subsection{Assessment of radio loudness} 
 
For all confirmed radio-detected objects their radio loudness was then estimated,
following Kellermann et al. (1989),
who define the  
radio index $R$ as
ratio of 6cm radio flux to optical flux at 4400\AA.
$R=10$ is commonly used to mark the ``border'' between radio-quiet and radio-loud objects,
originally calculated based upon the assumption
of similar spectral shapes in the optical
and radio band with $\alpha=-0.5$ (Kellermann et al. 1989).
Radio-surveys employed in the present study generally measure at 1.4 GHz.
Still keeping the same
assumptions, in particular, $\alpha=-0.5$,
we compute the radio index at 1.4 GHz as $R_{1.4}$=1.9$R$,
and we use $R_{1.4}$=19 as a reasonable distinction between radio-loud
and radio-quiet objects.
Optical blue magnitudes m$_{\rm B2}$ were extracted from the USNO-B1
catalogue and used to calculate $R_{1.4}$.
For comparison, blue magnitudes from the first epoch of
the USNO-B1 (m$_{\rm B1}$), the GSC2.2 (m$_{\rm Bj}$) and the USNO-A2 (m$_{\rm B}$)
catalogues were also
collected 
to get an impression on the uncertainty in
blue magnitude, which would be either due to source-intrinsic
variability and/or measurement uncertainties.  

As a last step, only a careful object-by-object check then allows to scrutinize
which objects are actually radio-loud NLS1 galaxies, or good candidates.
In particular, peculiar intrinsic spectral shapes, high redshift, or heavy
absorption, which will affect radio and optical band in a different way,
would invalidate the simple way of calculating 
the radio index $R$ as ratio of {\em observed} radio to optical flux density.
Also, merging galaxies with close double nuclei make counterpart
identification ambiguous. 
Finally, optical spectroscopic
classification of each radio-loud candidate
needs to be checked carefully in order to confirm 
its NLS1 nature. An uncertainty which  
generally cannot be overcome
with existing data given their non-simultaneity, 
is source variability either in the optical and/or radio band.

\subsection{Search results}

\subsubsection{General results}

Among the 128 NLS1s in the VQC  
we find that 35 have radio counterparts in one or more catalogues.
In Fig. 1 we present the distribution of their radio indices $R_{\rm 1.4}$.  
Formally, among this sample, $\sim$11 objects exceed   
or are close to the ``critical'' value of $R_{1.4}$=19 commonly used to 
distinguish between radio-quiet and radio-loud objects. 
The radio-loud NLS1s
are  
TEX11111+329, SDSSJ094857.3+002225, SDSSJ172206.03+565451.6, IRAS20181-2244, 
RXJ0134-4258, PKS0558-504, IRAS09426+1929,
SBS1517+520, RXJ16290+4007, 2E1346+2637, and RXJ23149+2243
(Tab. 1; sources are listed by their name as given in the
VQC). 
Among these,  we recover the previously known radio-loud
NLS1 galaxies PKS0558-504, RXJ0134-4258, SDSSJ172206.03+565451.6
and SDSS J094857.3+002225.{\footnote{Wang  
et al. (2004) claimed the detection of another radio-loud NLS1,
SDSSJ022119.8+005628.4. However, the position off-set 
between SDSS galaxy and the faint NVSS radio source
is 24$^{\prime\prime}$, much larger than the expected off-set 
of $\sim$7$^{\prime\prime}$ for faint sources.   
Furthermore, we do not find any radio counterpart in the FIRST
survey, indicating that the very faint NVSS counterpart, if real, 
is either very extended or variable. 
Deeper radio observations are needed to confirm the existence
of this radio source, and to see whether it indeed is the counterpart 
to the NLS1 galaxy.  }}  
PK2004-447, not included in our sample since it is a Seyfert rather than a quasar,
is unusual in almost lacking FeII emission in
its optical spectrum (Oshlack et al. 2001), which led to the suggestion
it could be a narrow-line radio galaxy or a type-2 AGN.  Given the large
scatter in FeII of the radio-loud objects of our sample, it is well possible
that PK2004-447 is similar, just at the low FeII end and we include it
in Tab. 1.

Based on average blue magnitudes of the non-detected sources and
given the NVSS flux limit ($\sim$2.5mJy), no more very radio-loud
sources are expected among the non-detections. Several more
borderline objects close to $R_{\rm 1.4} \approx 19$ could exist,
but only if their radio-flux was just exactly below the NVSS flux limit.  

Radio source positions agree well with the optical coordinates 
of the sources.  
FIRST counterpart position off-sets are typically $<$1$^{\prime\prime}$,
while NVSS off-sets are typically $<$5$^{\prime\prime}$. 
The radio emission of the radio-loud NLS1 galaxies is generally compact. 
Their FIRST extent is always $<$2$^{\prime\prime}$ which is consistent with
point sources (White et al. 1997). 

The 35 radio-detected sources cover about four orders of magnitude in 
radio index $R$. Below $\log R \simeq 1$ (Fig. 1) the data hint
at a rather smooth distribution of radio indices,
while only few objects are well above $\log R \simeq 1$.  
%
A number of our  objects (Fig. 2) fall into that part of the radio-loud regime sometimes
referred to as `radio-intermediate' ($\log R \simeq$ 1-2, 
as opposed to `truly radio-loud' above $\log R \simeq 2-3$;
e.g., Falcke et al. 1996), a regime
where objects are relatively rare, and which classically defined
the radio-loud radio-quiet bimodality of AGN.

\subsubsection{Variability}

Several sources were observed during both the NVSS and FIRST
survey, and these data can be used to assess the radio
variability of the galaxies. Generally, we find that NVSS and FIRST
radio fluxes agree well with each other. For three sources, fluxes 
differ by approximately a factor of 1.6--1.8. Among these,
variability of SDSSJ094857.3+002225 (integrated flux density 
 $S_{\rm FIRST} = 111.5$ mJy and $S_{\rm NVSS} = 69.5$ mJy)
is likely real
(Zhou et al. 2003). The other two sources are Mrk 478 
($S_{\rm FIRST} = 3.3$ mJy 
and $S_{\rm NVSS} = 5.2$ mJy)   
and PHL1811 ($S_{\rm FIRST} = 1.2$ mJy and $S_{\rm NVSS} = 2.1$ mJy). 
Higher NVSS than FIRST fluxes are actually expected in case of very extended
sources because part of the extended emission will be missed by FIRST
(e.g., Fig. 7 of White et al. 1997).
Indeed, we find indications that the radio emission of PHL1811 is extended
(major axis of 2.9$^{\prime\prime}$ reported in the FIRST catalogue,
and a value of $<74^{\prime\prime}$ in the NVSS catalogue).  
Inspecting the NVSS and FIRST images we find that Mrk 478 has
a nearby bright companion source which might have slightly contributed to
the NVSS radio flux estimate of Mrk 478.
Barvainis et al. (1996) report a dedicated radio observation of Mrk 478
at 1.49 GHz with $S = 4.46\pm{0.47}$ mJy, within the errors almost consistent 
with the NVSS and FIRST measurements.  

Among those sources detected in the radio band 
there are 10 in the FIRST data base which do not have a 
NVSS counterpart, while 7 in the NVSS data base do not have a FIRST
counterpart. The latter is just due to the non-availability of FIRST
observations at the respective coordinates.
The radio images of sources with FIRST-only detections were 
carefully inspected by eye
and it turns out that all FIRST detections are consistent within better
than a factor of two with the NVSS upper limits.

\subsubsection{Radio spectral indices}

Several sources were observed
 in the radio regime at more than one frequency. These data can be used to
estimate radio spectral shapes. 
Radio spectral indices $\alpha_{\rm r}$ are reported in Tab. 1.
$\alpha_{\rm r}$ shows a large scatter. 
Steep and flat spectrum sources
are observed. While the two galaxies SDSSJ094857.3+002225 and RXJ16290+4007
show very flat spectra ($\alpha_{\rm r}$=0.6 and 0.4, respectively) 
the majority of NLS1 galaxies are of steep spectrum type ($\alpha_{\rm r} < -0.5$).

\subsubsection{Notes on individual objects}

We comment here on the 11 apparently radio-loudest objects of this study, 
i.e.
those falling above or close to the critical radio index, $R_{1.4}=19$ (Fig. 1).
Emphasis is put on the question how safely these galaxies can be called ``radio-loud''
(Sect. 2.3) and how reliable their optical spectral classification is. 

{\bf TEX11111+329} is an ultraluminous infrared galaxy (ULIRG). Optical
spectroscopy was provided by Zheng et al. (2002) who classify
TEX11111+329 as NLS1 galaxy (FWHM$_{\rm H\beta} \simeq 1980$ km/s, [OIII]/H$\beta$=0.75,
FeII/H$\beta$=1.1). 
Lipari et al. (2003) report the detection of a high-velocity outflow component
in [OIII]5007 with $v \simeq 1300$ km/s. 
The galaxy is heavily absorbed, with at least
$E_{\rm B-V}$=1.08, estimated from  
the observed Balmer-line ratio (Zheng et al. 2002).
Assuming that this obscuration is not {\em intrinsic}
to the line-emitting clouds (the BLRs in NLS1s might be 
sufficiently distant from the nucleus that dust can survive) 
and 
correcting the observed optical continuum spectrum by      
A$_{\rm B}$=4.5 
leads to a much lower value of R$_{\rm 1.4} \simeq 20$. 
On the other hand, TEX11111+329 would also be classified as radio loud
just based on its radio power, which is $P_{\rm 1.4} \simeq 10^{25}$ W/Hz. 
The radio spectrum declines toward higher frequencies, and 
TEX11111+329 is not detected in an imaging survey at 15 GHz (Nagar et al. 2003).  
We keep the source as candidate radio-loud NLS1.

{\bf SDSSJ094857.3+002225} is a very radio-loud NLS1 galaxy (Zhou et al. 2003),
variable in the radio and likely the optical band. 
Inspecting the NVSS and FIRST radio image, we noticed a second
radio source very close to SDSSJ094857.3+002225,
FIRSTJ094901.5+002258, at a separation of 71$^{\prime\prime}$
(corresponding to a projected distance of 395 kpc) 
and extended by $\sim$5$^{\prime\prime}$. Suspecting it
to be either a knot in a radio jet of SDSSJ094857.3+002225, 
or a background source, we had a closer look
at optical images and cannot identify any optical counterpart.
Closest is a faint SDSS galaxy, SDSSJ094901.9+002306.4, at an off-set
of 10$^{\prime\prime}$, an unlikely counterpart of the FIRST radio source.    
It is possible that the radio source has a faint, so far undetected,
optical counterpart. 

{\bf SDSSJ172206.03+565451.6} is a very radio-loud NLS1 galaxy.
Depending on choice of blue magnitude, it may be comparably radio-loud
as SDSSJ094857.3+002225. Its radio loudness and optical spectral
properties are discussed in detail by Komossa et al. (2006).
In particular, its black hole mass in unusually small
given its radio loudness (Komossa et al. 2006).  
SDSSJ172206.03+565451.6 is the optically most variable among our sample. 
 
{\bf RXJ0134-4258} is a radio-loud NLS1 galaxy (Grupe et al. 2000)
with peculiar X-ray spectral variability 
(Komossa \& Meerschweinchen 2000, Grupe et al. 2000).
It was detected in the PMN survey and its radio-loudness
was confirmed by a dedicated follow-up observation by Grupe et al.  

{\bf IRAS\,20181-2244} was identified as a quasar with narrow emission lines by
Elizalde \& Steiner (1994). 
Halpern \& Moran (1998) rejected a ``type-2 quasar'' classification
of IRAS20181-2244 and called it a NLS1-type object, with FWHM$_{\rm H\beta}$=460 km/s
and the presence of broad wings in the Balmer lines.
Kay et al. (1999) found low-level continuum polarization (2\%) but no broad Balmer lines.
The X-ray emission of IRAS20181-2244 (e.g., Vaughan et al. 1999)
is rapidly variable (Halpern \& Moran 1998, Leighly 1999).
The Balmer decrement of the narrow lines implies significant reddening
(Tab. 1 of Halpern \& Moran 1998) but
the rapid X-ray variability argues against a full 
obscuration of the continuum source itself.
The {\sl ASCA} X-ray spectral fits indicate some 
excess absorption along the line of sight
and  Halpern \& Moran give $E_{\rm B-V} \approx 0.3$ mag as a reasonable
estimate of extinction intrinsic to IRAS\,20181-2244.
In the radio band, IRAS\,20181-2244 is also detected during the 
WISH (Westerbork in the Southern Hemisphere) survey at 352 MHz (de Breuck et al. 2002).  

{\bf RXJ16290+4007}  is a bright source in the X-ray (e.g., Giommi et al. 1991,
Elvis et al. 1992, Brinkmann \& Siebert 1994) and radio band 
(e.g., Giommi et al. 1991, Laurent-Muehleisen et al. 1997, White et al. 2000)
likely variable by $\sim$0.5 mag optically (Helfand et al. 1991). 
RXJ16290+4007 is known as flat spectrum radio quasar (FSRQ). Its FSRQ nature,
with respect to its X-ray and radio properties, is discussed in detail
by Padovani et al. (2002).  
Steady TeV gamma ray emission from RXJ16290+4007 was searched for, but not detected,
with the Whipple telescope (Falcone et al. 2004).
   Optical spectra provided by Bade et al. (1995) hinted at the NLS1 nature
of RXJ16290+4007, which was  
confirmed
by Schwope et al. (2000) and Grupe et al. (2004). 

 While Padovani et al. (2002) concentrated on the radio and X-ray
properties of RXJ16290+4007, interpreted its relatively steep X-ray
spectrum as due to synchrotron radiation, and suggested RXJ16290+4007
is the first member of a newly established class of HBL(high energy peaked
BL Lac)-like FSRQs, we rather follow the alternative idea that the
optical and X-ray properties of this source are driven by its ``narrow-line Seyfert 1
character''.

{\bf IRAS09426+1929} is ultraluminous in the infrared
with a single nucleus. 
Zheng et al. (1999) identify as counterpart
of the IR source a NLS1 galaxy at $z$=0.284, based on an optical
spectrum which shows weak [OIII], strong FeII and narrow Balmer lines
(no line parameters were measured).  
On the one hand, the host galaxy is detected as well 
and may contribute to the blue magnitude
which would make the computed value of the radio index $R$ 
a lower limit. On the other hand,
extinction measurements are not yet available for IRAS09426+1929. Since
the cores of ultraluminous infrared galaxies are often heavily obscured,
any calculation of $R$ based on the {\em observed} optical spectrum would
then make the value of $R$ an upper limit. This second effect is likely to exceed
the first.   

{\bf PKS0558-504} is a known radio-loud source 
(Remillard et al. 1986, Siebert et al. 1999), with an 
optical NLS1 spectrum (Remillard et al. 1986). 
It is highly variable in the X-ray band (e.g., Remillard et al. 1991,
Gliozzi et al. 2001, Wang et al. 2001). The rapid X-ray
variability and the lack of excess X-ray absorption
(O'Brien et al. 2001, Brinkmann et al. 2004) indicate an unobscured view on
the central engine.

{\bf RXJ23149+2243} was identified as quasar with strong FeII emission
by Wei et al. (1998). Xu et al. (2003) report emission-line measurements
which indicate a relatively large width of the Balmer lines. 
Our newly taken spectrum (Sect. 3) puts the galaxy at the border between NLS1 and Seyfert\,1,
depending on the way the width of H$\beta$ is measured. 
The [OIII]5007 line of this galaxy shows an exceptionally strong blue wing 
which is blueshifted from the narrow core by 1260 km/s (Sect. 3).  
RXJ23149+2243 is detected by IRAS and turns out to be a 
luminous infrared galaxy (LIRG). 

{\bf 2E1346+2637} is an ultrasoft AGN 
(photon index $\Gamma_{\rm x} \simeq -3.7$)
at high redshift, $z$=0.92 
(Puchnarewicz et al. 1994), optically classified as NLS1 galaxy 
(FWHM$_{\rm H\beta} \simeq$ 1900 km/s, weak [OIII] and presence of FeII; 
Puchnarewicz et al. 1992).  
The X-ray spectrum of the galaxy displays one of the 
strongest soft excesses known.   
Puchnarewicz et al. (1994) report a radio upper limit of 6 mJy at 4.85 GHz
and mention that this leaves a small margin that 2E1346+2637 is radio-loud.  
2E1346+2637 is detected by FIRST but not NVSS. The NVSS upper limit is
consistent with the FIRST measurement.    

{\bf SBS1517+520} is listed as NLS1 in the VQC 
but little else is known about this galaxy
(Bicay et al. 2000 reported its optical position).
Our analysis of the SDSS spectrum (Sect. 3) does not
confirm its classification as a NLS1 galaxy in the sense
that FWHM$_{\rm H\beta_{\rm g}}$ = 3220 km/s (based on single-component 
Gauss fit (Tab. 2); only the `direct' determination
of the width of $H\beta$, FWHM$_{\rm H\beta_{\rm d}}$ = 2030 km/s, 
puts it close to the border). Because other emission-line parameters share similarity
with those of NLS1 galaxies, we continue to include SBS1517+520 in our
tables.

\section{Optical spectroscopy of the radio-loud NLS1s}

In order to measure the optical 
emission-line and continuum properties of the radio-loud
NLS1s in a homogeneous way, we (re-)analyzed their optical 
spectra when available. The spectrum of RXJ0134-4258 was
taken by one of us (D.G.) at ESO's 1.5m telescope (Grupe 2004),
as was the spectrum of RXJ23149+2243 (D.X.),  
at the Xinglong 2.16m telescope (Xu et al. 2003), 
while those of SDSSJ094857.3+002225, SDSSJ172206.03+565451.6, RXJ16290+4007,
SBS1517+520 were extracted from  
the SDSS data base{\footnote{See http://cas.sdss.org/dr4/en/tools/explore/obj.asp}}. 
In addition, a new spectrum of RXJ23149+2243 was secured  
on November 25, 2005 at the 2.16m Xinglong telescope, motivated by the remarkable
structure of the [OIII]5007 profile of this galaxy.  
SDSS spectra of the other galaxies are not (yet) available.
For them, emission-line data were collected from the literature when published (Tab. 2).

The SDSS DR4 products 
were used.
These spectra are flux- and wavelength-calibrated by the spectroscopic
pipeline in the course of the DR4 processing.
The spectrum newly taken with the 600\,g\,mm$^{-1}$ grating
at the 2.16m Xinglong telescope was reduced
following standard procedures (see Xu et al. 2003). 
The IRAF package SPECFIT (Kriss 1994) was used for spectral analysis.

We first corrected the spectra for Galactic extinction,
using the Galactic E$_{B-V}$ in the direction of the
individual galaxies (Schlegel, Finkbeiner \& Davis 1998)
and a R=3.1 extinction law.  
We then fit a power law
to the underlying continuum using the ``continuum windows'' 
at 3010-3040, 3240-3270, 3790-3810,
4200-4230, 5080-5100, 5600-5630, 5970-6000, and 6005-6035~$\AA$ (Forster et al. 2001,
Vanden Berk et al. 2001) known to be relatively free from strong emission lines. 
The continua are generally well fit by single powerlaws
with indices between $\alpha = -1.3$ and $-2.3$ ($f_{\lambda} \propto \lambda^{+\alpha}$).

Since the optical-UV spectra of all sources show emission from FeII complexes, 
an FeII spectrum was then fitted and subtracted (excluding the
UV FeII multiplets around MgII$\lambda$2798 which are not included in the template)
using the optical FeII template and the technique described
by Boroson \& Green (1992). 

The FeII- and continuum-subtracted spectra were
then used to measure emission-line properties.
The emission lines were modeled by single Gaussian profiles or
combinations of Gaussians and Lorentzians. 
The approach was to fit single-component Gaussians to the 
weak forbidden lines, apply two-component Gaussians to [OIII]
when necessary, and compare double Gauss and Gauss-plus-Lorentz
profiles in their success in representing the H$\alpha$ and H$\beta$
line profile. In addition, a direct  measurement of the H$\beta$ line width
was performed (without any assumption on line profile
shape). H$\beta$ line results are marked with an index `g' or `d'
if referring to a Gaussian fit or the `direct' method, respectively,
and with `b', `n' or `totl' when referring to the broad, narrow or
whole emission component,
respectively, of a two-Gaussian decomposition of the line profile.  

Results for the diagnostically most important emission lines
are listed in Tab. 2. For the sake of simplicity and homogeneity,
the listed emission-line ratios are based on single
Gaussian fits to the lines, even though for some
sources this is not a perfect match of the line profile.  
The instrumental resolution was corrected for in the FWHMs measured by us. 
We confirm the NLS1 classification for almost all galaxies,
with the possible exceptions of SBS1517+520 and RXJ23149+2243 (see below).
Line widths of confirmed NLS1s range between FWHM$_{\rm H\beta_{\rm g}}$ = 930-1580 km/s,
while the flux ratio [OIII]/H$\beta_{\rm g}$ varies between 0.04 -- 1.1 
(results from Gauss fits are reported here). FeII is detected in all spectra
and ranges between FeII4570/H$\beta_{\rm g}$ = 0.5--3.2
(FeII4570/H$\beta_{\rm totl}$ = 0.5--2.4). The NLS1 galaxies
of our sample cover almost the whole range of FeII4570/H$\beta$
observed in the NLS1 population (e.g., Tab. 3
of Veron-Cetty et al. 2001), except that the lowest FeII/H$\beta$ ratios are missing in
our sample.   

SBS1517+520 and RXJ23149+2243 have rather broad Balmer lines,
and it depends on the way the profile of H$\beta$ is fit
whether or not they formally fullfil the criterion of
FWHM$_{\rm H\beta} < 2000$ km/s.  Both of them have 
FWHM$_{\rm H\beta_{\rm b}} > 2000$ km/s
while 
FWHM$_{\rm H\beta_{\rm d}} < 2000$ km/s. 
While the strict cut in FWHM is a useful working criterion for 
extraction of NLS1 galaxy candidates from larger samples,
it has been repeatedly noted that the cut value
is not well defined or even completely arbitrary.
Veron-Cetty et al. (2001) suggested that a better 
classification criterion than FWHM$_{\rm H\beta}$ may
actually be the FeII ratio, FeII4570/H$\beta_{\rm totl} > 0.5$.
According to this criterion, both SBS1517+520 and RXJ23149+2243
still classify as NLS1s.
After these cautious comments, we continue to keep the two
galaxies in the sample, even though they fail a strict
FWHM$_{\rm H\beta_{\rm b}} < 2000$ km/s criterion. 

RXJ23149+2243 is exceptional in showing a very strong and broad blue wing 
in the [OIII]5007 and [OIII]4959 lines. The presence of the blue wing was hinted
in the earlier low-resolution spectrum of Xu et al. (2003) and was confirmed
in the new spectrum we took in November 2005
(full results on this galaxy will be presented elsewhere). 
[OIII]5007 is well modeled by
two Gaussian lines. The core, with 
FWHM$_{\rm [OIII]_{\rm n}}$=620 km/s, is at the same redshift
as the peak of H$\beta$ while the wing, with FWHM$_{\rm [OIII]_{\rm b}}$=1560 km/s,
is highly blueshifted by 1260 km/s.     
SBS1517+520 also shows a blue wing in [OIII].  
The line is well fit by two Gaussians 
with FWHM$_{\rm [OIII]_{\rm n}}$=610 km/s
and FWHM$_{\rm [OIII]_{\rm b}}$=2070km/s, blueshifted by $v$=90 km/s.  
In the SDSS spectra of the other objects of our sample the 
blue wing is present, but not exceptionally
strong.  For SDSSJ172206.03+565451.6 the line blueshift is $v$=140 km/s,
for RXJ16290+4007 we measure $v$=230 km/s.     

\section{X-ray spectroscopy and variability}    

While some of the sources of our sample have no previous X-ray studies,
others (in particular, PKS0558-504) have been observed by
most major X-ray satellites. 
In order to obtain representative X-ray fluxes of all
our sources in a homogeneous way, including upper limits
for non-detected sources, we make use of the data obtained
with the PSPC (Positional Sensitive Proportional Counter) during 
the {\sl ROSAT} all-sky survey (RASS; Voges et al. 1999)
and the later phase of pointed observations.  

9 out of the 11 soures have X-ray detections, 8 of these in the RASS
and 3 in pointed observations. 
Countrates and hardness ratios were obtained for all
sources. 
Fixing the absorption at the Galactic values in the directions
of the individual sources,  hardness ratios were then converted to powerlaw
indices, and the absorption-corrected flux in the (0.1--2.4) keV band was 
determined. Powerlaw indices are in the range $\Gamma_{\rm x}=-1.9 .. -4.3$ (Tab. 2)
with the exception of the faint source SBS1517+520 ($\Gamma_{\rm x}=-1.1$,
likely caused by excess absorption). 
In case pointed observations
existed (RXJ16290+4007, 2E1346+2637, RXJ0134-4258) these were given preference
over the RASS data. 

For the two non-detected sources, TEX11111+329 and IRAS09426+1929,
upper limits were obtained.
However, we caution  
while soft X-ray non-detections convert into safe upper limits
for non-obscured sources, the same is not true for obscured sources.
This is likely the case for both TEX11111+329 and IRAS09426+1929. For completeness,
we report the 2$\sigma$ upper limits on the soft (0.1--2.4 keV) X-ray count rates in Tab. 2, 
but do not convert them into luminosities, and do not use them
for accretion rate estimates.  

We also inspected the RASS X-ray lightcurves of the X-ray brighter sources
and find that most are consistent
with constant source flux except PKS0558-504 which shows repeated flaring
activity. 
Variability of the X-ray fainter sources cannot be 
judged due to the poor statistics.    

\section{Discussion}

Judged on radio index, 
11 objects classify as radio-loud NLS1 galaxies
or candidates.  
Our study almost triples the number of recognized
radio-loud NLS1s if the candidates close to the
deviding line are confirmed. Implications of the new results
are discussed below.

\subsection{Comparison of the properties of the known radio-loud NLS1 galaxies}

Even though the number of known radio-loud NLS1s and the candidates
is still small, we have enough objects now to compare their X-ray,
optical and radio-properties, to have a first assessment of their
similarities or differences.
In particular, we checked whether the $\sim$11 radio-loud NLS1s 
are extreme in one or more of their multi-wavelength spectral
properties, i.e., whether they differ significantly 
from the radio-quiets in their properties,
and whether there is a trend {\em within} the sample
such that one or more properties correlate with radio loudness. 

Among optical emission lines, the ``first Eigenvector'' (Boroson \& Green 1992) is
evident among the objects of our sample in that weak [OIII]/H$\beta$ emission tends to go 
hand in hand with strong FeII emission and vice versa.
Both, [OIII]/H$\beta$ and FeII/$\beta$ vary strongly across the sample, almost
spanning the whole range observed in NLS1 galaxies.
In Fig. 3  we plot the location of the radio-loud NLS1s in the
FeII--[OIII] diagram, in comparison with one of the largest 
homogeneously analyzed optical NLS1 samples available to date (Xu et al. 2006). 
We do not find a strong correlation of any of the parameters
with radio loudness, but larger samples are needed to confirm this.

Even though radio loudness did not occur in the
sample of Marziani et al. (2001) for line widths 
FWHM$_{\rm H\beta_{\rm b}} < 4000$ km/s,
the radio-louds of our sample 
do populate similar regions as other AGN in the FWHM$_{\rm H\beta_{\rm b}}$-FeII
diagram (Fig. 4 of Marziani et al. 2001), and extend the range
where radio loudness is observed to significantly smaller FWHMs than
previously known (see also Sect. 5.5).

Given suggestions that the blue wings in the [OIII]5007 line of quasars are 
sometimes linked to interactions of the radio plasma with the NLR clouds 
(Leipski \& Bennert 2006, and references therein) we paid special
attention to the existence and strengths of blue wings in dependence
of radio loudness but do not find any correlation, even though 2-3 out of
the 11 sources do show very intense blue wings, two with strong blueshifts,
among the highest values ever observed. 

Most NLS1s of our sample have steep  
radio spectra (Tab. 1) in the frequency band 0.33--1.4--4.85 GHz{\footnote{The non-simultaneity
of the data, taken with different instrumental
configurations,  has to be kept
in mind, but unlikely all scatter in $\alpha_{\rm r}$ among the sample can be traced
back to variability, else variability in individual cases would be huge.}}
and their radio emission is compact.
As such, they share some similarity with `compact steep spectrum sources'
(CSS, Peacock \& Wall 1982) which are believed to be young 
radio sources which did not yet evolve strongly 
(Marecki et al. 2005, and references therein). 
SDSSJ094857.3+002225 and RXJ16290+4007 have inverted radio spectra. 
Only one source, SDSSJ094857.3+002225 (Zhou et al. 2003), shows
significant variability in the radio-band
by a factor $\sim$1.6.

Soft X-ray spectra are generally steep, but come with large errors.
The mean {\sl ROSAT} powerlaw index of the radio-loud NLS1s
of our sample{\footnote{note that a number of them were first identified by
X-ray observations}} (excluding SBS1517+520 which has unusually flat index), 
$\Gamma_{\rm x}=-2.7$, is similar to that of
soft X-ray selected NLS1 galaxies ($\Gamma_{\rm x}=-2.96\pm{0.41}$, Grupe et al. 2004).     
PKS0558-504 (e.g., Remillard et al. 1991), IRAS20181-2244 (Halpern \& Moran 1998)  
and RXJ16290+4007 (Padovani et al. 2002) are highly variable in 
the X-ray band 
while RXJ0134--4258
strongly changed its spectral shape
(Grupe et al. 2000, Komossa \& Meerschweinchen 2000).

\subsection{Starburst contribution ?}  

It was suggested that NLS1s are AGN in an early phase of evolution.   
This includes the possibility
of enhanced starburst activity in these objects 
(e.g., Mathur 2000, Shemmer et al. 2004).
Could  starbursts contribute to or dominate the radio power 
of the radio-loud NLS1s of our sample? 
Generally,
normal and starburst galaxies are much less luminous than quasars 
in the radio regime.
In particular, all of our NLS1s also exceed 
the radio powers of a sample of the most radio-luminous
starbursts, all of them in luminous and ultraluminous
infrared galaxies (Smith et al. 1998), 
$ \log P_{\rm 4.85,SB} \simeq 22.3-23.4$ W/Hz. 
Radio powers place all of our NLS1s with $R_{1.4} > 19$
in the radio-loud regime ($\log P > 24$,
Joly 1991), some of them close to the lower border,
though.

At first glance, a significant starburst contribution might be suspected for the
two ultraluminous infrared galaxies among our sample, 
TEX11111+329 and IRAS09426+1929. 
We used the tight correlation between IR luminosity
and radio luminosity of starburst galaxies (Yun et al. 2001, and references therein)
to predict the expected starburst radio power, given the 60$\mu$ IR
luminosities of these two galaxies measured by IRAS. 
According to equ. (4) of Yun et al. (2001), with
$\log P_{\rm 1.4,NVSS} \simeq 25$ W/Hz 
TEX11111+329 exceeds by a factor of 10 the contribution
expected from a starburst, while for IRAS09426+1929
the measured radio power,
$\log P_{\rm 1.4,NVSS} \simeq 24.7$ W/Hz, is a factor 5 above
the predicted starburst contribution.  
TEX11111+329 has radio observations at several frequencies. 
The spectral index is relatively steep, but comparable
to other sources of our sample. 
The spectral index between 0.33-1.4 GHz is
similar to SDSSJ172206.03+565451.6 and PKS0558-504
while towards higher frequencies it is
comparable to RXJ0134-4258. 

The same comparison between observed radio power and expected starburst
contribution was done for the three other galaxies with IRAS 60$\mu$ flux
measurements. We find that the observed radio powers of PKS0558-504,
IRAS20181-2244, and RXJ23149+2243 exceed by factors of 115, 6, and 6, respectively,
the predicted starburst contribution.  

We conclude that radio-luminous starbursts are unlikely to dominate
the radio emission of most of our sources.

\subsection{Black hole masses}

Reverberation mapping of Seyfert galaxies established a relation
between the radius of the broad line region and the optical luminosity
(Kaspi et al. 2005, Peterson et al. 2004, and references therein). 
Assuming that the broad line region (BLR) clouds are
virialized (e.g., Wandel et al. 1999), the
black hole mass can then be estimated as
$M_{\rm BH} = G^{-1}\,R_{\rm BLR}\,v^{2}$. 
The BLR radius, $R_{\rm BLR}$,
in dependence of the optical luminosity at 5100\AA~
is given by 
$R_{\rm BLR} = 22.3[{{\lambda\,L_{\lambda}{\rm(5100\AA)}}\over{10^{44}{\rm erg/s}}}]^{0.69}$ ld
(equ. (2) of Kaspi et al. 2005). 
The velocity $v$ of the BLR clouds is usually estimated
from the FWHM as $v = f\,{\rm FWHM}$ where $f={\sqrt{3}\over{2}}$ for
an isotropic cloud distribution.
While the relation is now well established for nearby Seyfert galaxies,
few NLS1s have been reverberation mapped so far (Peterson et al. 2000). 
Following standard practice, we assume equ. (2) of Kaspi et al. (2005) 
is indeed applicable to NLS1 galaxies as well and proceed in
estimating the BH masses of our sample. 

$L_{\lambda}{\rm(5100\AA)}$ was derived from the SDSS 
spectra (corrected for Galactic extinction) 
when available, or else estimated from blue magnitudes $m_{\rm B2}$ of the
USNO-B1 catalogue (note that some of the galaxies of our
sample vary in the optical band which introduces an uncertainty
in the black hole mass estimates. Most objects, however, vary
by less than a factor 2).
The FWHM of the broad component from the two-component
fit of H$\beta$ was used for the black hole mass estimates  
reported in Tab. 3 when optical spectra were available.
Else, results on FWHM from single-component line fits
collected from the literature (Tab. 2) were used for
an order-of-magnitude estimate.  

We obtain black hole masses
in the range  $M_{\rm BH} \simeq (2-10)\,10^{7}$ M$_{\odot}$
(objects with two-component decomposition of H$\beta$ available)
and $M_{\rm BH} \simeq (0.2-9)\,10^{7}$ M$_{\odot}$ 
(objects with single component fits to H$\beta$; this underestimates
FWHM$_{\rm H\beta_{\rm b}}$ and therefore underestimates the black hole mass) 
for our sample. These masses are surprisingly small,
given the radio loudness of the galaxies (Fig. 4).
In fact, all galaxies  
fall right into an unpopulated
regime of the ``Laor diagram'' (Fig. 2 of Laor 2000)
which plots radio loudness in dependence of black hole
mass.
While larger samples filled up some originally
empty areas in the Laor diagram, and while some discussion
has emerged whether there is any clear dependence of black hole
mass on radio loudness at all (e.g., Lacy et al. 2001,
Oshlack et al. 2002, Woo and Urry 2002, McLure \& Jarvis 2004,
Best et al. 2005),
our radio-loud NLS1 galaxies are still located in an area
very sparsely populated by galaxies. 
Firstly, because the black hole masses are unusually small
given the radio loudness of the galaxies of our sample (e.g., Fig. 2 of Laor
et al. 2000, Fig. 3 of Lacy et al. 2001, Fig. 2 of McLure \& Jarvis 2004,
Fig. 10 of Metcalf \& Magliocchetti 2006).
Secondly, because most radio indices $R$ of our objects are in
a range which is classically known to be less populated 
by quasars and which defined the radio-loud radio-quiet bimodality
of AGN (with radio-quiet quasars below $R \approx 1-10$ 
and many radio-loud quasar above $R \approx 100-1000$ (e.g, Fig. 10
of Woo \& Urry 2002)).  
An even more extreme outlier is PKS2004-447
which shows lower BH mass and higher radio index $R$ than objects
of our sample (Oshlack et al. 2001).

While of unusually low mass when compared to radio-loud quasars,
{\em among the NLS1 population itself} our radio-loud NLS1s are
actually at the {\em high-mass end} (for comparison, in the sample
of Grupe \& Mathur (2004) NLS1 masses range between 6\,10$^5$--4\,10$^7$ M$_{\odot}$
with only one object around 10$^8$ M$_{\odot}$).  
To some extent, the lack of higher mass BHs among
NLS1s in general may just be a `selection' effect which
arises from the standard definition of NLS1s as objects with
FWHM$_{\rm H\beta} < 2000$ km/s.  Since there is a clear correlation
between H$\beta$ luminosity and H$\beta$ width (e.g., Fig. 15
of Veron-Cetty et al. 2001), higher luminosity AGN  
should have broader Balmer lines finally exceeding
FWHM $>$ 2000 km/s and thus would no longer be called
NLS1s. Ideally, any NLS1 defining
criteria should incorporate that effect (Wills et al. 2000).  

In any case, while we cannot exclude that NLS1s with higher BH masses
do exist, the fact that the radio-loud objects of our sample
cluster in a previously almost unpopulated parameter space of
$M_{\rm BH}$ and $R$ is still remarkable.  
We caution there are indications that 2-3 galaxies of the sample 
are actually beamed sources (Sect. 5.7.1). Any contribution
of beamed emission to the optical band makes our estimate
of black hole mass an {\em upper limit}, while, on the other hand,
we {\em underestimate} the black hole mass if the BLR clouds are 
arranged in a disk rather than a sphere (Vestergaard et al. 2000).
Among the sources of our sample, SDSSJ172206.03+565451.6 is the radio
loudest with `reliable' radio index measurement, since no
indications for beaming and likely no absorption in the optical band
(Komossa et al. 2006).   

\subsection{Accretion rate}

With knowledge of BH mass and X-ray luminosity it is possible to perform an  
estimate of the accretion luminosity relative to the Eddington
value.
We intentionally use the X-ray luminosity here, and we do not
apply any bolometric correction, since it is unlikely constant for
all NLS1 galaxies, and since, in particular, the EUV part of their SED
is unknown{\footnote{There have been suggestions that the X-rays connect in one
`big blue bump' to the optical band on the one hand, 
but findings that NLS1s are underluminous
in the UV on the other hand, and inspections of individual objects lead to
expectations of flat EUV-SEDs in some cases, soft-excesses in other cases. Ideally,
bolometric corrections would therefore be done for each object individually, based
on a measurement of the multi-wavelength SED on the one hand, and detailed 
emission-line modeling on the other hand in order to constrain the EUV-part of
the SED from emission-line ratios and H$\beta$ line flux. }}.

We find that 
the ratio of (0.1-2.4 keV) luminosity to Eddington luminosity (Tab. 2)  
ranges between  
$L_{\rm x}$/$L_{\rm Edd}$ = 0.02 (RXJ23149+2243) and 0.7 (RXJ16290+4007)
{\em without} any bolometric correction which
might typically be an additional factor 5--10
(Tab. 14 and 15 of Elvis et al. 1994).
An exception is SBS1517+520 with $L_{\rm x}$/$L_{\rm Edd}$ = 0.007.
We suspect that this very faint X-ray source is absorbed in X-rays 
by a (dust-free) absorber which would also explain its very flat powerlaw with index
$\Gamma_{\rm x} \simeq -1.1$ which is formally obtained when fixing the absorption
at the Galactic value.  
We excluded TEX11111+329 and IRAS09426+1929 from this estimate
since no meaningful X-ray upper limit could be obtained.  

For comparison with $L_{\rm bol}$/$L_{\rm Edd}$ values reported in the literature
we also estimated bolometric luminosities based on the optical luminosity
at 5100\AA, and a constant bolometric correction factor of 9 (e.g., Sect. 2 of Warner et al. 2004,
Greene et al. 2006); i.e. 
$L_{\rm bol} = 9 \times \lambda L_{\lambda}$(5100\AA).     
This gives an average $L_{\rm bol}$/$L_{\rm Edd}$=1.2 
(again excluding TEX11111+329 and IRAS09426+1929, plus IRAS20181-2244)
which is higher than the NLS1 average of Warner et al. (2004; 0.67, but
note that they used CIV1549 rather than H$\beta$ for BH mass estimates) 
but not as high as observed in the most 
luminous quasars (e.g., Warner et al. 2004, Shemmer
et al. 2004).

\subsection{Radio-FeII connection, comparison with Population A/B sources, 
 and the radio-loud radio-quiet bimodality in `Eigenvector 1'}

Among {\em broad-line} quasars
the nature of the FeII emission has been  much discussed
in the literature with respect to radio properties (e.g., Joly 1991,
Miller et al. 1993).
Two models were considered to explain the connection between
radio and FeII emission in radio-loud sources: an orientation-dependent model
of FeII emission (e.g. Jackson \& Browne 1991), and
a jet model in which FeII is produced when the jet interacts
with the ambient medium (Joly 1991).
Boroson \& Green (1992) and Boroson (2002) find that 
radio loudness is statistically linked to
the physical parameters that drive ``Eigenvector 1'' in the sense
that quasars with strong [OIII] and weak FeII are preferentially radio-loud
while strong FeII emitters are generally radio-quiet objects
(for an example of an exception, see Zhou et al. 2002).

The radio-loud  NLS1s of our sample do not fit in this scheme in the sense
that their FeII emission does not show a clear dependence on 
radio loudness, but almost covers the whole observed
range in NLS1 galaxies (Fig. 3; cf. Fig. 4
 of Marziani et al. 2001).
In particular, RXJ0134-4258, the $\sim$4th radio-loudest
of our sample,  is one of the strongest FeII
emitters known.

Sulentic et al. (2003) demonstrate that the radio-loud radio-quiet
distinction of AGN is obvious in their `Eigenvector 1' which links the width
of the broad component of H$\beta$ with the equivalent-width 
ratio of FeII to H$\beta$ (their Fig. 3), in the sense that radio-loud and radio-quiet objects
occupy different domain spaces. NLS1 galaxies are not among their sample.
Sulentic et al.  define two populations of AGN according to their emission-line
properties. `Population A' are sources with FWHM$_{\rm H\beta_{\rm b}} < 4000$ km/s,
while FWHM$_{\rm H\beta_{\rm b}} > 4000$ km/s in  `population B' sources.
In their sample, radio loudness is almost exclusively confined to population B. 
The radio-loud NLS1 galaxies of our sample fall into the rarely populated
regime of Fig. 3 of Sulentic et al., most of them in the `gap' between
radio-louds and radio-quiets, but not to an extent that the bimodality
is obliterated.

\subsection{Inclination}
Orientation plays an important role in AGN unification scenarios 
of both Seyfert galaxies (Antonucci 1993, Elvis 2000) and 
radio galaxies (Urry \& Padovani 1995).
Inclination also enters black hole
mass and accretion rate estimates and consequently
affects black hole mass -- bulge mass
relations of NLS1s and their cosmological implications (see e.g. the discussion
in Collin \& Kawaguchi 2004).

The orientation of NLS1s w.r.t our line of sight has repeatedly been addressed,
and arguments and evidence in favor of and against a pole-on view have been
presented.
Originally, for instance, a pole-on view
was suggested to explain the small width of the Balmer lines
(e.g., Osterbrock and Pogge 1985, Puchnarewicz et al. 1992,
Bian \& Zhao 2004).
The disk emission of FeII is also anisotropic 
and FeII emission is strongest when the disk is viewed  face-on
(e.g., Miller et al. 1993, Marziani et al. 01).
A pole-on orientation was also suggested for individual NLS1s with
extreme variability (e.g.,    
PKS0558-504: Remillard et al. 1991).
Correcting for these orientation effects, systematic or random in nature,
is also important in the context of black hole mass estimates
and black hole mass -- bulge mass relations of NLS1s, which
make use of Balmer line widths.
On the other hand, Smith et al. (2004) argued that, based on their
polarization properties, there are no indications that NLS1s are preferentially
viewed face-on. 

Radio observations of NLS1 galaxies allow us to re-address orientation issues.
Among our sample,
there are 2-3 NLS1s which are likely beamed (SDSSJ094857.3+002225, RXJ16290+4007,
and possibly PKS0558-504; Sect. 5.8.1), implying 
a pole-on view onto the `central engine'. 
If the width of the Balmer lines was dominated by orientation, we would then expect 
these objects to show very narrow Balmer lines.  
This is not the case (FWHM$_{\rm H\beta_{\rm b}}=1710$ and 1800 km/s, respectively
for SDSSJ094857.3+002225 and RXJ16290+4007), though.  Larger samples of beamed NLS1
galaxies are needed to put this on a firmer statistical basis.

\subsection{Frequency of radio-loud NLS1 galaxies}

Of the 128 NLS1 galaxies in the Veron quasar catalogue,
90\% are located within the NVSS survey area. Among these, 
7\% are formally radio-loud ($R_{1.4} {\mathrel{\hbox{\rlap{\lower.55ex \hbox {$\sim$}}
        \kern-.3em \raise.4ex \hbox{$>$}}}} 19$), while only
2.5\% exceed $R=100$.
One or a few of these sources are most likely significantly
absorbed and one or two are at
the border between Seyfert\,1 and NLS1 galaxy.
Given a fraction
of $\sim$13-20\% of radio-louds among optically selected quasars
(e.g., Visnovsky et al. 1992, LaFranca et al. 1994, Falcke et al. 1996), 
radio-loud NLS1s are more rare
than radio-loud quasars.

In order to compare directly the fraction of radio-louds among narrow-line 
and broad line type I quasars   
listed in the VQC,  we have also extracted the broad line quasars
from the VQC, cross-correlated them with
the NVSS catalogue, and determined the fraction of radio-louds the
same way we did for the NLS1 galaxies.    
We find that $\sim$90\% of the objects are located within the NVSS survey area.
Among these, $\sim$20\% are radio-loud 
($R_{1.4} {\mathrel{\hbox{\rlap{\lower.55ex \hbox {$\sim$}}
    \kern-.3em \raise.4ex \hbox{$>$}}}} 19$), 
and $\sim$14\% exceed $R=100$.
Again, this indicates a lack of radio-louds among NLS1s, particularly 
at high values of radio index $R$.

When comparing the frequency of radio-loud NLS1s with radio-loud type I quasars,
two things have to be kept in mind.
Firstly, the mere definition of NLS1s as objects with FWHM$_{\rm H\beta} < 2000$ km/s
likely biases the observed black hole mass distribution towards excluding the most massive
black holes. Since there are indications that radio-loud AGN have, on average,
higher BH masses (e.g., Metcalf \& Magliochetti et al. 2006), we may automatically
miss some of the most massive radio-loud NLS1s because they escape the definition
as NLS1s and are rather called
Seyfert 1s.

Secondly, specific to the present sample, great attention was paid to
each single object in that cautious comments were issued in case of evidence for
excess extinction, and in that line profiles were fit with a number of different models.
The same detailed object-by-object analysis should be applied to comparison
samples of quasars, when citing the fraction of radio-loud quasars in comparison
with that of radio-loud NLS1 galaxies.

Keeping this in mind, it still appears that radio-loud NLS1s are less abundant
than radio-loud quasars, and we proceed in attempting to explain this distinction.

The cause of radio loudness in quasars is still unknown, 
even though a number of different models have been suggested and
investigated including the influence of spin (e.g., Wilson \& Colbert 1995, 
Blandford 2000, Ye \& Wang 2005),
black hole mass (e.g., Laor 2000, Metcalf \& Magliocchetti 2006),
the rate of cooling gas in the host galaxy (Best et al. 2005) and 
beaming in radio-intermediate FSRQs which were suggested to be relativistically
boosted radio-quiets (Miller et al. 1993, Falcke et al. 1996).  

Two key questions regarding the radio properties of the NLS1 galaxies are:
Why is the phenomenon of
radio loudness rarer in NLS1 galaxies than in quasars?
And, related to that: Is the mechanism which drives the radio loudness the same
in both types of objects, or is a fundamentally different
mechanism at work in the few radio-loud NLS1s?
We discuss several models in turn.

\subsubsection{Beaming}  

We first explore the possibility that 
NLS1s might be radio-quiet but intrinsically beamed sources.
In this scenario, one then expects that other optical and X-ray properties of 
the radio-loud NLS1s
are affected as well. 
For instance, FeII emission might be extreme  
if it originates in a disk which is viewed face on
(e.g., Miller et al. 1993). 
Also, X-ray soft excess  emission is expected to be stronger if the sources
are viewed face on for the case of a geometrically thick accretion disk (Madau 1998). 
At the same time there might be a flatter non-thermal contribution to X-rays
from the jet emission. 

While a few  sources of our sample are indeed good candidates
for beaming, others are not.
RXJ16290+4007 and SDSSJ094857.3+002225 are likely beamed, but only RXJ16290+4007
is radio-`intermediate'.
RXJ16290+4007 likely is a blazar of FSRQ type (e.g., Padovani et al. 2002)
and SDSSJ094857.3+002225 exhibits strong variability 
in the radio band (Zhou et al. 2003).  
These are the only two sources among our sample 
with very flat radio spectra (Tab. 1). 
Apart from that, PKS0558-504 exhibits unusual X-ray flaring activity
and there have been repeated speculations that this may be
related to beaming effects in a jet (Remillard et al. 1991, Gliozzi
et al. 2001, Wang et al. 2001).  An origin in the accretion disk
is an alternative, though, and this
latter model is preferred by Wang et al. (2001) and 
by Brinkmann et al. (2004) to explain the XMM data. 

Other sources of our sample are steep spectrum radio sources where
strong beaming is generally not expected. 
X-ray spectra are steep (even those of the two FSRQs), 
arguing against the dominance of a non-thermal Doppler-enhanced 
contribution to the soft X-ray spectrum. Beamed sources more typically
show $\Gamma_{\rm x} \simeq -1.5 .. -1.6$.

\subsubsection{Accretion mode}

While estimated Eddington accretion rates  
are in line with other findings that NLS1 galaxies
are accreting close to or above their Eddington limit,
the radio loudness of such sources is unexpected at first
glance, given observations that radio emission is suppressed
in accretion high-states in some types of Galactic and
extragalactic sources. 
According to Maccarone et al. (2003, and references therein),
Galactic X-ray binaries in soft/high-state and AGN show quenched 
radio emission for relatively high accretion rates.
Even though the analogy between Galactic black holes and NLS1 galaxies
has been drawn repeatedly (e.g., Pounds et al. 1995), it is still unclear how
far it reaches.
Greene et al. (2006, and references therein) more generally suggested that
radio loudness is anti-correlated with $L_{\rm bol}/L_{\rm Edd}$ (their Fig. 3)
in a sample of Seyferts and PG quasars, albeit with quite some scatter. 
Whether this trend still holds for high-redshift,
highly accreting quasars has yet to be explored.                                   
So far, these observations suggest, whatever the 
mechanism to suppress radio emission
for high $L_{\rm bol}/L_{\rm Edd}$ could also be responsible 
for the lower fraction of radio-louds among NLS1s.

While it is generally agreed upon that jet formation (and thus radio
emission) is intimately linked to the presence and structure of an
accretion disk, the actual mechanism of launching a jet from the
inner disk regions is not yet really understood 
(see, e.g., Celotti \& Blandford 2001 and Meier 2003 for recent reviews).
One may speculate that the observed differences originate from
differences in the accretion mode between stellar mass and
super-massive black hole accretors and between different
classes: It is generally agreeed upon that at relatively low
accretion rates (in terms of the Eddington rate), radiatively
inefficient accretion flows prevail (e.g., Beckert \& Duschl 2002).
Independent of yet not fully understood details of these flows, they
seem to be potentially efficient in producing radio emission
themselves (e.g., Yi \& Boughn 1998, Narayan et al. 1998). At rates
closer to, but still well below the Eddington limit disks turn
optically thick and geometrically thin
(see Wang et al. 2002 for
a suggestion how radio-emission might be enhanced in this case).
Barring details at these low accretion rates the properties of the flows seem to be rather
similar between stellar mass and super-massive black hole accretors.

When the accretion rate reaches or even surpasses the Eddington
rate, however, the physical state of the accretion process becomes
less clear. One has to keep in mind that the Eddington rate has been
derived for a perfectly spherically symmetric situation. The
situation in accretion disks, however is rather different from such
an ideal case. While at rates close to the (spherical) Eddington
limit, accretion disks in all likelihood are no longer geometrically
thin ({\it slim disks\/}, see, e.g., Collin \& Kawaguchi 2004),
their deviation from sphericity is even more important. This can
lead to simple geometrical effects allowing for super-Eddington
accretion rates (Heinzeller and Duschl, {\it in prep.\/}), but also
to instabilities in the structure of the disks, for instance the
photon bubble instability as discussed by Begelman (2002), and in
all likelihood to combinations of such geometrical and physical
effects (e.g., Okuda et al. 2005).

An additional potentially relevant difference between stellar mass
and super-massive black hole accretors may be the mass of the disk.
While in the stellar case the masses of the accretion disk are
certainly well below that of the accreting body, this is
considerably less clear for the AGN case, and may be even different
for different AGN types. In the innermost disk regions, selfgravity
certainly plays no role. Already at moderately large radii, however,
this may change. With such changes, long-term modifications of the
mass supply rate for the inner regions comes along. Different levels
of selfgravity in the disks may very well lead to differences in the
accretion process.

Therefore, different accretion modes have the potential to explain
the observed differences in radio loudness between Galactic and AGN
sources on the one side, and among the different AGN classes on the
other side. Before one, however, can unambiguously identify whether,
and if so, which aspect of the accretion mode gives rise to the
differences, a much better understanding is required of how these
different accretion modes relate to jet formation.


\subsubsection{Black hole spin}

Another parameter which could account for radio loudness
but at the same time leave optical emission-line parameters and correlations
mostly unaffected, is black hole spin. 
It is expected to have a strong direct influence on radio jet emission
if the radio jet is  powered by the extraction of black hole rotational energy
(Blandford \& Znajek 1977), or may indirectly be relevant if the
jet is powered by the rotational energy of the inner disk 
(Blandford \& Payne 1982).   

If, indeed, rapidly spinning black holes
are responsible for radio loudness of NLS1s and quasars, 
why then would NLS1s {\em on average} harbor {\em less} rapidly spinning
black holes than Seyferts (to account for the lower frequency
of radio-loudness among NLS1s)?
Both BH-BH mergers and accretion
will affect BH spin (e.g., Merritt 2002, Hughes \& Blandford 2003,
Volonteri et al. 2005). 
It has been suggested that NLS1 galaxies are less evolved than Seyferts
in that their black holes are still in the process of growing
by rapid accretion (e.g., Mathur 2001, Grupe \& Mathur 2004, Botte et al. 2004,  
Mathur \& Grupe 2005). 
Since accretion processes are known to be very
efficient in spinning up the accreting black holes (Volonteri et al. 2005,
and references therein) one then indeed expects that 
the final stages, `normal' Seyferts and quasars, should 
have, on average, more rapidly spinning black holes than the NLS1s.

At the same time, spin cannot be the only quantity
making objects radio-loud, else all quasars must be
more radio-loud than NLS1s.  
Thus, spin would be a necessary but not sufficient condition.  
If, indeed, spin plays a role, then the few radio-loud NLS1s 
should be those with the more rapidly spinning BHs, already
more evolved than others. One may thus expect that they
lie closer to the black-hole mass -- velocity dispersion  ($M_{\rm BH} - \sigma$) 
relation of Seyfert galaxies
than other NLS1s.  
We have plotted the objects of our sample in the $M_{\rm BH} - \sigma$ 
plane of Mathur \& Grupe (2004; their Fig. 1) and indeed find
that they are systematically closer to, or on, the $M_{\rm BH} - \sigma$  
relation followed by Seyferts. 

Future X-ray observations of the iron line may help
to search for signs of more rapidly spinning BHs in the radio-loud NLS1s,
but less rapidly spinning BHs in the NLS1 population as a whole.  

In summary, while we cannot fully exclude beaming to operate in all sources,
the source properties are generally such that this mechanism is not favored
(with the three exceptions mentioned above).
Accretion mode, possibly in combination with spin, is a possibility.

\subsection{Future work}

The results of this study lead us to identify a number of
important follow-up observations:

High quality optical spectra of the 11 radio-loudest objects of our sample
do not yet exist, or are not yet published. Spectra of high resolution
and S/N would allow a search for similarities/differences in
emission lines between
radio-loud and radio-quiet NLS1s the same way it is now rigorously done for 
broad- and narrow-line objects (e.g., Veron et al. 2001, 
Sulentic et al. 2002, Bachev et al. 2004).
Especially important is a reliable decomposition of the faint Balmer lines
like H$\gamma$ and H$\delta$ into broad and narrow component, and deblending
from [OIII]4363, such that extinction measurements become possible 
for those objects which have H$\alpha$ redshifted out of the observable 
optical band.  
This knowledge is essential for   
the assessment of radio loudness (in reddened objects,
radio loudness is overpredicted), and for black hole mass calculation. 

Good optical spectra of the radio-loud and almost radio-loud
NLS1s, and a comparison sample of quasars, will enable BH mass 
and bulge mass estimates using
H$\beta$ and [OIII] or low-ionization lines, 
and ideally also absorption features  (e.g., Grupe \& Mathur 2004, Botte et al. 2004,
Greene \& Ho 2005)
which will then allow
us to test more rigorously whether claimed systematic trends of higher black hole
masses in radio-loud vs. radio-quiet quasars extend   
to the NLS1 population.  

X-ray observations of all sources are needed in order to
measure accurate spectral shapes, few of them
available so far, and to determine the amount of absorption.
Deeper X-ray observations are required to
measure iron-line profiles in search for (less) rapidly spinning black holes
in NLS1 galaxies.

How much of the scatter in radio spectral index of the NLS1s  
is real and how much is caused by the non-simultaneity of the data, 
should be tested with simultaneous radio observations at different frequencies
and monitoring for variability. The latter will also constrain the number
of beamed sources. 
Radio observations of higher spatial resolution
should be carried out in order to confirm the 
compactness of the NLS1s of our sample and thus also to check
how far the similarity with known CSS sources goes. 

Larger samples of AGN with a `relaxed' NLS1 classification criterion
where the cut in FWHM is dependent on luminosity will enable us to address
the question whether the fraction of radio-loud NLS1s systematically
changes (increases) with FWHM$_{\rm H\beta}$ and whether (radio-loud) NLS1s
with higher BH masses emerge. The largest sample to date of NLS1s
from the SDSS data base, but still using an FWHM cut-off, is
presently being compiled by Zhou et al. (priv. comm.).    

Apart from future multi-wavelength follow-up observations of the small
sample of radio-loud NLS1s presented here,
Virtual Observatory tools will become more and more
important to study trends throughout the whole NLS1 and AGN population,
since they will allow the extraction of multi-wavelength information from existing data bases
in a homogeneous way 
which then enables us to study bulk properties of populations on the one hand,
and to identify exceptional  
individual sources  on the other hand. 

{\sl Comparison with the work of Whalen et al. (2006)}: While
our paper was being refereed, a paper by Whalen et al. (2006) appeared
on the astro-ph preprint server. These authors present optical follow-up spectroscopy
of FIRST radio sources identified as NLS1 galaxies, including 16 candidate
radio-loud NLS1s ($>$50\% of their sample has radio powers $<10^{24}$ W/Hz). 
Their findings are consistent with ours in that both works do not find 
strong correlations between optical line parameters and radio-loudness,
and black holes masses show similar mass ranges. Being a radio-selected 
sample, the fraction of radio-loud NLS1 galaxies is higher in
the sample of Whalen et al.; on the other
hand a number of the sources of Whalen et al. are below a radio power of $10^{24}$ W/Hz, 
and  we note that 
sources significantly above a radio index $R$=1000 are missing in both studies.

\section{Summary and conclusions}

We have presented a study of the radio properties of NLS1 galaxies,
with emphasis on the search for radio-loud NLS1s. 
The results of this study can be summarized as follows:

11 candidate radio-loud NLS1 galaxies were identified,
significantly increasing the number of known radio-loud NLS1s. These are 
radio loud judged on both, radio index and radio power.  
Radio-detected NLS1s cover several orders of magnitude in radio index $R$.
The distribution is rather smooth below $R \approx 10$.

The fraction of detected radio-loud NLS1s is $\sim$ 7\%, smaller than
the fraction of radio-louds among quasars. Only 2-3\% of the NLS1s
exceed a radio index $R > 100$ (SDSSJ172206.03+565451.6, SDSSJ094857.3+002225
and possibly the absorbed source TEX11111+329).
This could partly, but likely not fully,
be a `selection effect' arising from the NLS1 definition
as objects with FWHM$_{\rm H\beta} < 2000$ km/s independent
of luminosity.

The radio-loud NLS1s are generally steep spectrum sources and compact.
Two have inverted spectra.  
NVSS and FIRST radio fluxes
are consistent with each other in most cases, indicating little variability.    

Accretion rate estimates show that most sources likely
accrete close to or above the Eddington limit.
Black hole masses are generally at the upper end observed
in NLS1 galaxies ($M_{\rm BH} \approx (2-10)\,10^{7}$ M$_{\odot}$,
for objects with optical spectra of good quality), 
but still small when compared to some samples
of radio-loud quasars. In particular, the NLS1s of our sample  
fill up a previously scarcely populated area of $M_{\rm BH}-R$ diagrams.

Optical properties of the radio-loud NLS1s
are similar to the NLS1 population as a whole. 
In particular, their FeII and [OIII] emission covers almost the whole range
observed in NLS1 galaxies. Their radio properties, however,
extend the range of radio-loud objects to those with small FWHM$_{\rm H\beta}$.  
The fraction of [OIII] `blue-winglers' is high.  X-ray properties are consistent
with the NLS1 population as a whole, with $\Gamma_{\rm sx} \simeq -2 .. -3.5$ and 
$L_{\rm sx} \simeq 10^{44-46}$ erg/s; some sources highly variable, others constant
throughout the observation interval. 

The starburst contribution to the radio emission, estimated
for those sources with IRAS 60$\mu$ detections, is  
small or completely negligible. 
The AGN is expected to dominate any starburst in
radio emission by factors 5--115.  

We do not find indications for beaming in most of
the sources, 
arguing against the explanation that we see
intrinsically radio quiet but beamed sources. 
Accretion mode and spin might be factors to account for
the lower frequency of radio-loud NLS1 galaxies than quasars.   
 
Future studies of radio-loud NLS1s and follow-up observations
of the sample presented here  will provide
important information on distinguishing between different NLS1 models, 
on (the existence of) the radio-loud radio-quiet dichotomy of
AGN, and on accretion disk - jet models.

\acknowledgments

GAVO was funded by the Bundesministerium f\"ur Bildung und Forschung (BMBF)
 under contract no. 05\,AE2EE1/4.
DX acknowledges the support of the Chinese
 National Science Foundation (NSF)
 under grant NSFC-10503005.
This work is partially based on a spectrum taken at the Chinese
2.16m Xinglong telescope. We thank Jiaming Ai for taking the Xinglong spectrum 
of RXJ23149+2243,  
Hongyan Zhao for sharing part of his observing time with us,
and Jian-Min Wang for discussions. We thank Ed Moran for his referee report.   
This research made use of the SDSS and {\sl ROSAT} archives, 
the NED and VizieR service,
 and the following catalogues: IRAS, USNO-A2, USNO-B1, GSC2.2,
 FIRST, NVSS, SUMSS, WENSS, PMN, 87GB, PKS, and the Catalogue of Quasars and Active Nuclei.

\clearpage


\begin{figure}
\plotone{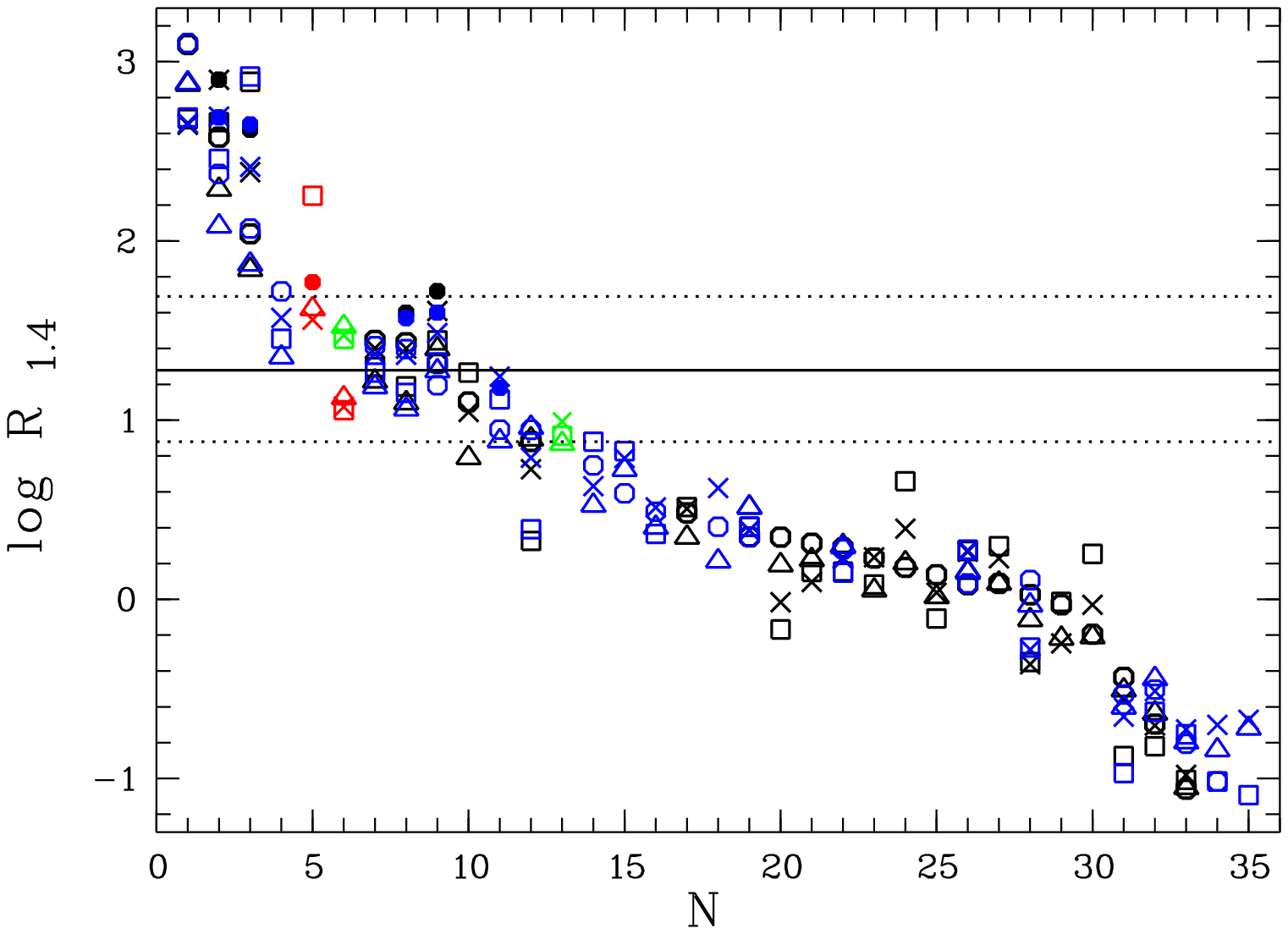} 
\caption{\small Radio-detected NLS1 galaxies selected from
the VQC, ordered
according to radio loudness. 
The solid line marks the ``deviding
line'' between radio-loud and radio-quiet objects, $R_{1.4} = 19$ (corresponding
to $R = 10$; see text for definitions). Calculation of $R_{1.4}$ 
is generally based on USNO-B1 blue magnitudes m$_{\rm B2}$ (squares)
and FIRST (plotted in black) and/or NVSS (plotted in blue) radio detections,
except RXJ\,0134-4258 (PMN data at 4.85 GHz, plotted in red), PKS0558-504 (PMN data, red; and 
SUMSS data at 0.8 GHz, green), and RXJ22179-5941 (SUMSS data, green).  
Several FIRST/NVSS objects have
radio detections at other wavelengths. These are not overplotted here, but
are used for estimates of radio spectral indices (Tab. 1). 
In order to have
an estimate on the uncertainty in m$_{\rm B}$ - either due to real source variability
or measurement uncertainties - we also show $R_{1.4}$ calculated
using m$_{\rm B1}$ (circles) of the USNO-B1 catalogue, m$_{\rm B}$ of the
USNO-A2 catalogue (triangles), m$_{\rm B_J}$ taken 
from the GSC2.2 catalogue (crosses), and corrected 4400\AA~fluxes obtained 
from the SDSS, ESO and Xinglong spectra (small filled circles).  
The dotted line indicates a change in blue magnitude 
by $\Delta m_{\rm B}$ = $\pm$1\,mag.  
{\scriptsize{Plotted ratios are for the following 
galaxies, from left to right: N=1: TEX11111+329,
2 = SDSSJ094857.3+002225, 3 = SDSSJ172206.03+565451.6, 
4 = IRAS20181-2244, 5 = RXJ0134-4258, 6 = PKS0558-504, 7 = IRAS09426+1929,
8 = SBS1517+520, 9 = RXJ16290+4007, 10 = 2E1346+2637, 11 = RXJ23149+2243,
12 = IRAS11598-0112, 13 = RXJ22179-5941, 14 = IRAS00275-2859, 15 = IRAS20520-2329,
16 = RXJ01354-0426, 17 = MS12510-0031, 18 = IRAS11058+7159, 19 = HS0710+3825,
20 = SBS1152+523, 21 = SBS1126+516, 22 = IRAS13349+2438, 23 = RXJ15308+2026,
24 = RXJ16196+2543, 25 = RXJ11425+2503, 26 = RXJ15475+1024, 27 = FBS1002+437,
28 = PG1404+226, 29 = RXJ17025+3247, 30 = RXJ12257+2055, 31 = PG1448+273,
32 = MRK478, 33 = PHL1811, 34 = IZw1, 35 = IRAS13224-3809.  }}
}
\end{figure}
\normalsize

\clearpage
\begin{figure}
\plotone{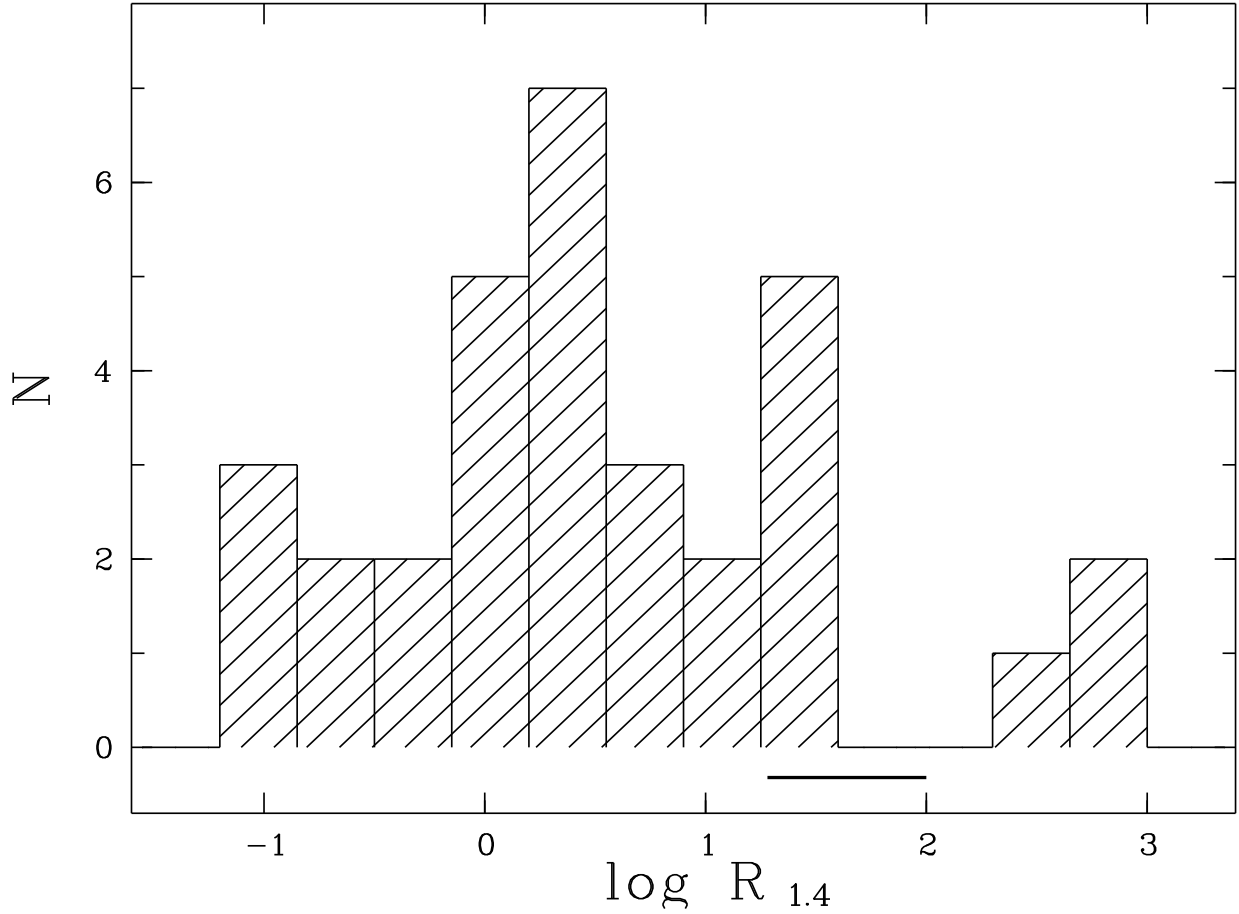}
\caption{Distribution of the radio index $R_{1.4}$ of the
NLS1 galaxies of our sample. $R_{1.4}$ was calculated using
NVSS radio fluxes and USNO-B1 m$_{\rm B2}$ magnitudes, except
if only FIRST fluxes and/or m$_{\rm B1}$ magnitudes were available
which were then used. The horizontal fat bar marks the range
within the radio-loud regime which is sometimes referred to as 
`radio intermediate'.  
}
\end{figure}

\clearpage

\begin{figure} 
\plotone{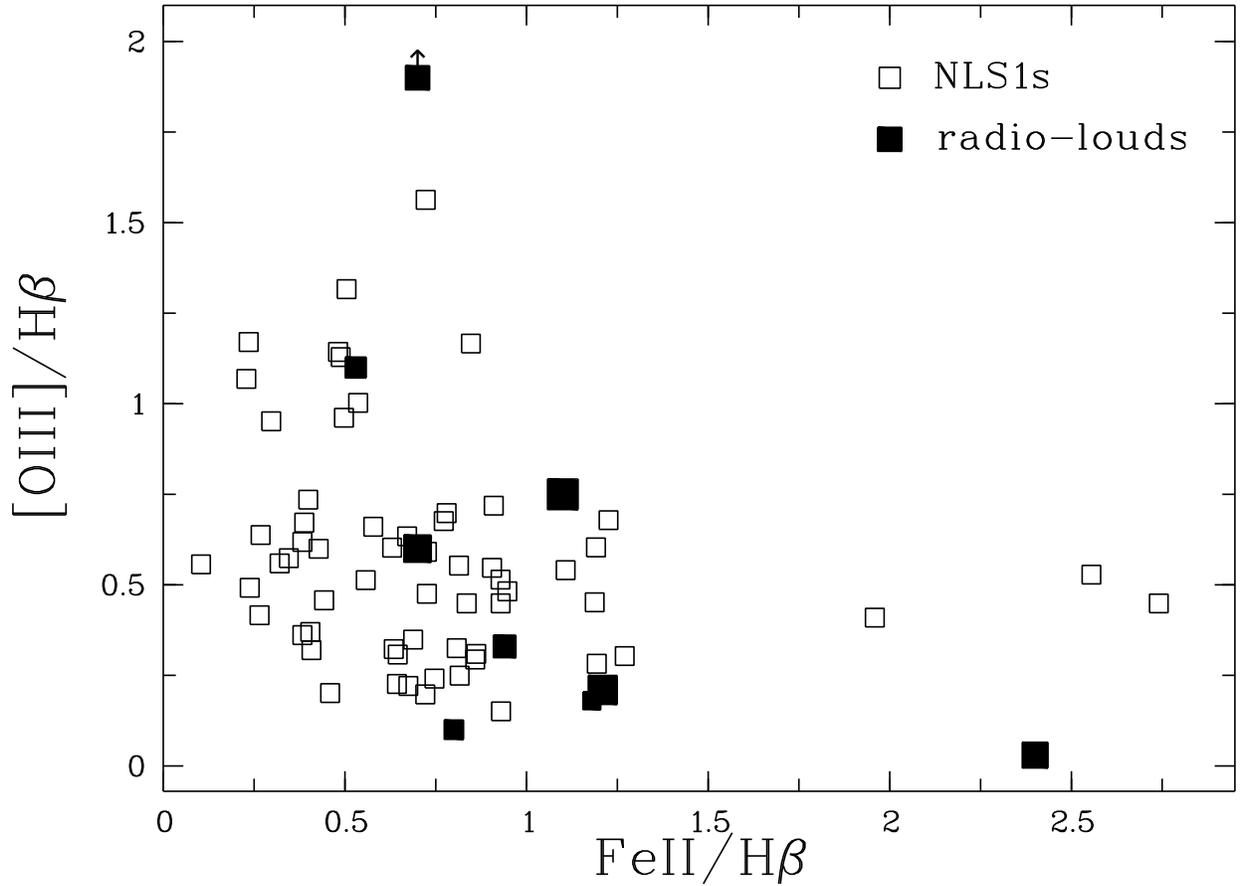} 
\caption{Distribution of the radio-loud NLS1 galaxies of our sample  (filled squares)
in the FeII/H$\beta_{\rm totl}$-[OIII]/H$\beta_{\rm totl}$ diagram of NLS1 galaxies (open squares; Xu et al. 2006). 
Square size codes radio-loudness; the larger the square diameter the radio-louder the source.  
The source IRAS20181-2243 is off the plot, indicated by the arrow. 
}
\end{figure}

\clearpage 

\begin{figure} 
\plotone{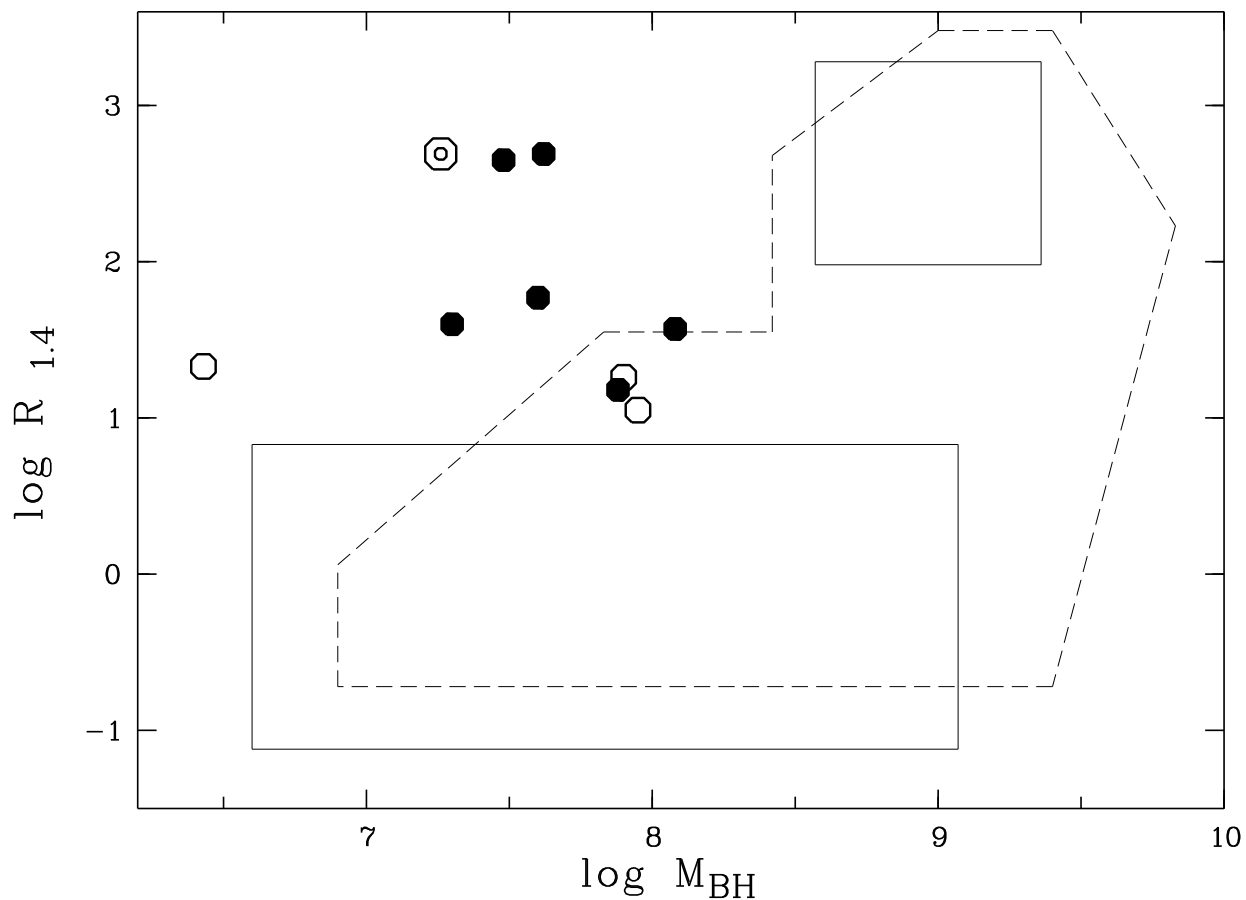}
\caption{The dependence of radio-loudness on black holes masses
for the radio-loud NLS1 galaxies
of our sample for which spectra were available to us (fat circles) and based on
spectra in the literature (open circles). The dashed and solid lines mark areas 
populated by the bulk of the radio sources of the samples of Laor (2000; solid line)
and Lacy et al. (2001, dashed line). 
The open double-circle marks TEX11111+329 which is possibly strongly absorbed.  
See text for further details, including comments
on individual sources. 
If both FRIST and NVSS radio observations were available, 
radio indices shown in this plot were calculated based on the NVSS data.  }

\end{figure}

\clearpage

\begin{table}
\rotate
\caption{Summary of the radio properties of radio-loud NLS1s and 
candidates,  
plus PKS\,2004-447 (a peculiar NLS1, Oshlack et al. 2001). 
The range in radio index $R$ of each galaxy is defined by the choice of
the blue magnitudes from different epochs/catalogues as in Fig. 1,
and is based on FIRST observations if available.  
$z$ is the redshift, $\nu_{\rm R}$ gives the observed frequency at which $R$ was calculated,
$\alpha_{\rm r}$ is the radio powerlaw spectral index ($f_{\rm \nu} \propto \nu^{+ \alpha}$), 
and $\nu_{1}$,$\nu_{2}$ specify the frequency interval in which $\alpha_{\rm r}$
was calculated.
}
\begin{tabular}{lcccrcl}
\tableline
\tableline
name & $z$ & $R$ & $\nu_{\rm R}$ & $\alpha_{\rm r}$ & $\nu_1$,$\nu_2$ & comments, references  \\
     &   &   &    GHz        &                  &     GHz         &        \\
\tableline
TEX11111+329 & 0.189 & 1249..445 & 1.4 & $-0.56$ & 0.33,1.4 & optically absorbed \\  
             &       &           &     & $-1.24$ & 1.4,4.85 &   \\
             &       &           &     & $< -2.07$ &  1.4,15 & upper limit at 15 GHz ([1]) \\
SDSSJ094857+002225 & 0.584 & 793..194 & 1.4 & & &  \\
                 & & $>$1000 & 5 & 0.59 & 2.7,4.85 & [2], variable  \\
SDSSJ172206+565451 & 0.425 & 773..70 & 1.4 & $-0.69$ & 0.33,1.4 &  [4] \\
RXJ0134-4258  & 0.237 & 178..36 & 4.85 & $-1.43$ & 4.85,8.4 & see also [3] \\
IRAS20181-2244 & 0.185   &  52..22   & 1.4 & $-0.51$ & 0.35,1.4 &  \\
RXJ16290+4007 & 0.272 & 41..21 & 1.4 & 0.42 & 1.4,4.85 & \\
              &       & 68..35 & 4.85 &     &          & \\
IRAS09426+1929 & 0.284 & 28..16 & 1.4 & & &  \\
PKS0558-504 & 0.137 & 16..14 & 4.85 & $-0.45$ & 2.7,4.85 &  \\
            &       &        &      & $-0.38$ & 0.84,2.7 & \\ 
RXJ23149+2243 & 0.168 & 18..8   & 1.4 & & &  \\
2E1346+2637 & 0.915 & 18..6 & 1.4 & & &  \\
SBS1517+520 & 0.371 & 27..12 & 1.4 & & &  \\
\tableline
PKS\,2004-447 & 0.24 & 6320..1710 & 4.85 & $-0.67$ & ATCA & [5], variable  \\
\tableline
\tableline
\end{tabular}

\mbox {    } \\
References: [1]: Nagar et al. (2003), [2]: Zhou et al. (2003), [3]: Grupe et al. 2000,
 [4]: Komossa et al. 2005, [5]: Oshlack et al. 2001 
\end{table}

\clearpage

\begin{table}
\rotate
\caption{Summary of optical and X-ray properties of the radio-loud NLS1s 
of our sample.
w$_{\rm H\beta}$ and w$_{\rm [OIII]}$ correspond to
FWHM$_{\rm H\beta}$ and FWHM$_{\rm [OIII]5007}$, respectively, in km/s.
The three entries for FWHM$_{\rm H\beta}$ are based on a single-component
Gauss fit, a direct fit, and a two-component fit with narrow
component fixed to the same width as [OIII]5007, respectively.
[OIII]5007, FeII4570 and H$\alpha$ are intensity ratios relative to H$\beta_{\rm g}$.
`CR' and $\Gamma_{\rm x}$ refer to the {\sl ROSAT} PSPC countrate in cts/s and the
soft X-ray photon index, respectively. 
Column `ref' provides the references for the optical spectroscopy. 
Entries of the first six objects are based on our (re-)analysis of SDSS
and other spectra. The other entries were taken from the literature. In the latter case,
multi-component fits to the Balmer lines are generally not available,
and only results from single Gaussian fits to the emission lines are listed.  
}
\small  
\begin{tabular}{lcccccccl}
\tableline
\tableline
name & w$_{\rm H\beta}$ &  w$_{\rm [OIII]}$ & [OIII] & FeII & H$\alpha$ & CR & $\Gamma_{\rm x}$ & ref \\
\tableline
SDSSJ094857+002225  & 1420/~960/1710+w$_{\rm [OII]}^{1}$ & 400$^{2}$ & 0.2 & 1.3 & - & 0.04 & $-$2.2 & [1]\\
SDSSJ172206+565451  & 1580/1490/1990+w$_{\rm [OIII]}$ & 490 & 0.6 & 0.7 & - & 0.08 & $-2..-3$ & [1],[6] \\ 
RXJ0134-4258     & ~930/1040/1860+w$_{\rm [OIII]}$ & 510 & 0.04 & 3.2 & 3.4 & 0.20 & $-$2.2$^{4}$ & [1] \\ 
RXJ16290+4007    & 1260/1290/1800+w$_{\rm [OIII]}$ & 310 & 0.4 & 1.1 & 3.3 &  0.83 & $-$3.0 & [1] \\  
RXJ23149+2243     & 1670/1630/2680+w$_{\rm [OIII]}$ & 620$^{3}$ & 0.2 & 1.4 & - & 0.11 & $-$1.9  & [1]  \\
SBS1517+520      & 3220/2030/3810+w$_{\rm [OIII]}$ & 1140 & 1.1 & 0.5 & - & 0.015 & $-$1.1 & [1] \\   
\tableline
TEX11111+329     & 1980 & 1480 & 0.75 & 1.1 &  & $<$0.04 & - & [2] \\
IRAS09426+1929   & $<$2000 & - & - & - & - & $<$ 0.03 & - & \\
PKS0558-504      & 1500 & - & 0.1 & 1.56$^{*}$ & -  & 4.20 & $-$2.8 & [3]\\
2E1346+2637      & 1840 & 1130$^{2}$ & - & 1.0 &  - & 0.08  & $-$3.4 & [4] \\
IRAS\,20181-2244 & 460 & 695 & 3.4 & 0.7 & 6.7 & 0.04 & $-$2.6 & [5]\\
\tableline
\tableline
\end{tabular}

\normalsize
$^{1}$Width of narrow component fixed to [OII]3227 since [OIII]5007 is too weak;
$^{2}$the reported FWHM is that of [OII] rather than [OIII] (in case of 2E1346+2637 the [OII] width
is not de-convolved from the instrumental profile).   
$^{3}$The [OIII]profile has a very strong and broad blue wing. The FWHM reported here is
that of the narrow core of the line only. 
$^{4}$During the RASS, the spectrum was significantly steeper ($\Gamma_{\rm x}=-4.3$). 

$^{*}$Sum of both FeII complexes.

References: [1]: this paper, [2]: Zheng et al. 2002, [3]: Remillard et al. 1986,
[4]: Puchnarewicz et al. 1996, [5]: Halpern \& Moran 1998, [6]: Komossa et al. 2005 
\end{table}
\normalsize

\clearpage 

\begin{table}
\rotate
\caption{
Estimated black hole masses and Eddington ratios
$L_{\rm x}$/$L_{\rm Edd}$ (multiplied by a factor 10) for NLS1s with SDSS or other spectra   
(broad component of H$\beta$ used for BH mass estimate;
first 6 objects) 
and other NLS1s
(single Gauss fits to H$\beta$ from
literature used for BH mass estimates). }  
\small
\begin{tabular}{lccl}
\tableline
\tableline
name & $M_{\rm BH}$ & 10$\times$$L_{\rm x}$/$L_{\rm Edd}$ & comments  \\
     & M$_{\odot}$        &                      &          \\
\tableline
SDSSJ094857+002225 & 4\,10$^7$ & 4.5 & source might be beamed; see also [1] \\
SDSSJ172206+565451 & 3\,10$^7$ & 3 & \\
RXJ0134-4258 & 4\,10$^7$ & 1 & \\
RXJ16290+4007 & 2\,10$^7$ & 7 & source might be beamed  \\
SBS1517+520 & 1\,10$^8$ & 0.07 & likely X-ray absorbed $\rightarrow$ $L_{\rm x}$/$L_{\rm Edd}$ is lower limit \\
RXJ2314+2243 & 8\,10$^7$ & 0.2 &  \\
\tableline
PKS0558-504 & 9\,10$^7$ & 7 & see also [2]  \\ 
TEX11111+329 & 2\,10$^7$ & - & likely optically absorbed \\
IRAS09426+1929 & $<9\,10^7$ & - & FWHM$_{\rm H\beta}$$<$2000 km/s assumed; probably optically absorbed \\
2E1346+2637 & 8\,10$^7$ & 4 & \\
IRAS\,20181-2244 & 3\,10$^6$ & 6 & \\
\tableline
\tableline
\end{tabular}
[1]: Zhou et al. (2003), [2]: Wang et al. 2001
\normalsize

\end{table}

\end{document}